%% file: main.tex
\documentclass[12pt,a4paper]{scrartcl}
\pdfoutput=1

\usepackage[utf8]{inputenc}
\usepackage[T1]{fontenc}
\usepackage{amsmath,amssymb,mathtools}
\usepackage[separate-uncertainty=true,retain-unity-mantissa=false]{siunitx}
\usepackage{slashed}
\usepackage{physics}
\usepackage{graphicx}
\usepackage{booktabs,multicol,multirow}
\usepackage[shortlabels]{enumitem}
\usepackage{hyperref}
\usepackage[capitalize]{cleveref}
\usepackage{xspace}
\usepackage[noblocks]{authblk}
\usepackage[dvipsnames]{xcolor}
\usepackage{soul}
\usepackage[textsize=tiny]{todonotes}
\usepackage{subcaption}
\usepackage{ulem}
\usepackage{bbold}
\usepackage{comment}

\usepackage[sorting=none,%
  citestyle=numeric-comp,%
  bibstyle=mynumeric,%
  giveninits=true]{biblatex}
\input{abbr.tex}

\addtokomafont{disposition}{\rmfamily}

\title{\Large New tools for dissecting the general 2HDM}
\author[1]{Henning Bahl\footnote{\href{mailto:hbahl@uchicago.edu}{hbahl@uchicago.edu}}}
\author[1,2,3]{Marcela Carena\footnote{\href{mailto:carena@fnal.gov}{carena@fnal.gov}}}
\author[1]{Nina M. Coyle\footnote{\href{mailto:ninac@uchicago.edu}{ninac@uchicago.edu}}}
\author[1]{\\Aurora~Ireland\footnote{\href{mailto:anireland@uchicago.edu}{anireland@uchicago.edu}}}
\author[1,3,4]{Carlos E.M. Wagner\footnote{\href{mailto:cwagner@uchicago.edu}{cwagner@uchicago.edu}}}
\affil[1]{Department of Physics and Enrico Fermi Institute, University of Chicago, 5720~South~Ellis~Avenue, Chicago, IL~60637~USA}
\affil[2]{Fermi National Accelerator Laboratory, P.O. Box 500, Batavia, Illinois, 60510, USA}
\affil[3]{Kavli Institute for Cosmological Physics, 5640 South Ellis Avenue, University of Chicago, Chicago, IL 60637}
\affil[4]{HEP Division, Argonne National Laboratory, 9700 Cass Ave., Argonne, IL 60439, USA}
\date{}
\titlehead{\hfill \parbox{5cm}{\vspace{-2cm}\begin{flushright} \texttt{EFI 22-8} \\ \texttt{FERMILAB-PUB-22-740-T} \end{flushright}}}

\begin{document}
\maketitle

\begin{abstract}\noindent
    \input{sec_abstract.tex}
\end{abstract}
\setcounter{footnote}{0}

\newpage

\tableofcontents

\newpage


\section{Introduction}
\label{sec:intro}

\input{sec_intro.tex}


\section{The general 2HDM}
\label{sec:2HDM}

\input{sec_2HDM.tex}


\section{Methods for bounding matrix eigenvalues}
\label{sec:math}

\input{sec_math.tex}


\section{Perturbative unitarity}
\label{sec:unitarity}

\input{sec_unitarity.tex}


\section{Boundedness from below}
\label{sec:BFB}

\input{sec_BFB.tex}


\section{Vacuum stability}
\label{sec:stability}

\input{sec_stability.tex}


\section{\cp Violation in the general 2HDM}
\label{sec:pheno}

\input{sec_CP.tex}


\section{Conclusions}
\label{sec:conclusions}

\input{sec_conclusions.tex}


\section*{Acknowledgments}
\sloppy{ C.W. would like to thank H. Haber and P. Ferreira for useful discussions and comments. 
Fermilab  is  operated  by  Fermi  Research  Alliance, LLC under contract number DE-AC02-07CH11359 with the United States Department of Energy. M.C.\ and C.W.\ would like to thank the Aspen Center for Physics, which is supported by National Science Foundation grant No.~PHY-1607611, where part of this work has been done. C.W.\ has been partially supported by the U.S.~Department of Energy under contracts No.\ DEAC02- 06CH11357 at Argonne National Laboratory. The work of C.W.\ and N.C.\ at the University of Chicago has also been supported by the DOE grant DE-SC0013642. H.B.\ acknowledges support by the Alexander von Humboldt foundation. A.I.\ is supported  by the U.S. Department of Energy, Office of Science, Office of Workforce Development for Teachers and Scientists, Office of Science Graduate Student Research (SCGSR) program, which is administered by the Oak Ridge Institute for Science and Education (ORISE) for the DOE. ORISE is managed by ORAU under contract number DE-SC0014664.}


\newpage
\appendix
\label{sec:appendix}

\input{appendix.tex}


\clearpage
\printbibliography

\end{document}

%% file: abbr.tex
\renewcommand{\cp}{\ensuremath{{\cal CP}}\xspace}

\newcommand{\gev}{\;\text{GeV}\xspace}

\renewcommand{\order}[1]{\ensuremath{{\cal O}(#1)}}

\NewBibliographyString{refname}
\NewBibliographyString{refsname}
\DefineBibliographyStrings{english}{%
  refname = {Ref\adddot},
  refsname = {Refs\adddot}
}
\DeclareCiteCommand{\ccite}
  {%
  \ifnum\thecitetotal=1
    \bibstring{refname}%
  \else%
    \bibstring{refsname}%
  \fi%
  \addnbspace\bibopenbracket%
  \usebibmacro{cite:init}%
   \usebibmacro{prenote}}
  {\usebibmacro{citeindex}%
   \usebibmacro{cite:comp}}
  {}
  {\usebibmacro{cite:dump}%
   \usebibmacro{postnote}%
   \bibclosebracket}
\newrobustcmd*{\Ccite}{\bibsentence\ccite}

%% file: sec_abstract.tex
Two Higgs doublet models (2HDM) provide the low energy effective theory (EFT) description in many well motivated extensions of the Standard Model. It is therefore relevant to study their properties, as well as the theoretical constraints on these models. In this article we concentrate on three relevant requirements for the validity of the 2HDM framework, namely the perturbative unitarity bounds, the bounded from below constraints, and the vacuum stability constraints. In this study, we concentrate on the most general renormalizable version of the 2HDM --- without imposing any parity symmetry, which may be violated in many UV extensions. We derive novel analytical expressions that generalize those previously obtained in more restrictive scenarios to the most general case. We also discuss the phenomenological implications of these bounds, focusing on \cp violation.

%% file: sec_intro.tex
The Standard Model (SM)~\cite{ParticleDataGroup:2018ovx} relies on the introduction of a Higgs doublet, whose vacuum expectation value breaks the electroweak symmetry~\cite{Englert:1964et,Higgs:1964pj,Guralnik:1964eu}. This mechanism generates masses for the weak gauge bosons and charged fermions, as well as potentially the neutrinos (although there may be other mass sources for the latter). The Standard Model Higgs sector is the simplest way of implementing the Higgs mechanism for generating the masses of the known elementary particles. However, it is not the only possibility, and may be easily extended to the case of more than one Higgs doublet without violating any of the important properties of the SM. Moreover, one of the simplest of these extensions --- two Higgs doublet models (2HDMs)~\cite{Branco:2011iw}  --- appears as a low-energy effective theory of many well motivated extensions of the Standard Model (e.g.\ those based on supersymmetry~\cite{Haber:1993an,Pilaftsis:1999qt,Carena:1995bx,Lee:2015uza,Carena:2015uoe,Bahl:2018jom,Bahl:2019ago,Murphy:2019qpm,Bahl:2020mjy,Bahl:2020jaq} or little Higgs~\cite{Low:2002ws}). 

Two Higgs doublet models may differ in the mechanism of generation of fermion masses. If both Higgs doublets couple to fermions of a given charge, their couplings will be associated to two different, complex sets of Yukawa couplings, which would form two different matrices in flavor space. The fermion mass matrices would be the sum of these, each multiplied by the corresponding Higgs vacuum expectation value. So diagonalization of the fermion mass matrices does not lead to the diagonalization of the fermion Yukawa matrices. Such theories are then associated with large flavor violating couplings of the Higgs bosons at low energies --- a situation which is experimentally strongly disfavored. Hence, it is usually assumed that each charged fermion species couples only to one of the two Higgs doublets. In most works related to 2HDM, this is accomplished by implementing a suitable $\mathbb{Z}_2$ symmetry. The different possible charge assignments for this $\mathbb{Z}_2$ symmetry then fix the Higgs--fermion coupling choices and define different types of 2HDMs.

This $\mathbb{Z}_2$ symmetry not only fixes the Higgs--fermion couplings but also forbids certain terms in the Higgs potential that are far less problematic with respect to flavor violation. As a starting point for an investigation of the phenomenological implications of these terms, we will in this work discuss the theoretical bounds on the boson sector of the theory (without any need to specify the nature of the Higgs-fermion couplings).  We will concentrate on the constraints that come from the perturbative unitarity of the theory, the stability of the physical vacuum, and the requirement that the effective potential is bounded from below. Existing works~\cite{Klimenko:1984qx,Kanemura:1993hm,Akeroyd:2000wc,Ferreira:2004yd,Horejsi:2005da,Ginzburg:2005dt,Barroso:2005sm,Ivanov:2006yq,Maniatis:2006fs,Barroso:2007rr,Ivanov:2007de,Ferreira:2009jb,Branco:2011iw,Jurciukonis:2018skr} focus either on the $\mathbb{Z}_2$-symmetric case or only provide a numerical procedure to test these constraints in the general 2HDM (see \ccite{Song:2022gsz} for a recent work on analytic conditions for boundedness-from-below). We will go beyond current studies by deriving analytic bounds that apply to the most general, renormalizable realization of 2HDMs. Our conditions will be given in terms of the mass parameters and dimensionless couplings of the 2HDM tree-level potential. At the quantum level, however, these parameters are scale dependent; although we will refrain from doing so here, one can apply these conditions at arbitrarily high energy scales by using the renormalization group evolution of these parameters.  

Our article is organized as follows. In \cref{sec:2HDM} we introduce the scalar sector of the most general 2HDM that defines the framework for most of the work presented in this article. \cref{sec:math} reviews three theorems from linear algebra which will allow us to derive analytic bounds in the coming sections. In \cref{sec:unitarity}, we concentrate on the requirement of perturbative unitarity. \cref{sec:BFB} presents the bounds coming from the requirement that the tree-level potential be bounded from below. In \cref{sec:stability}, we discuss the vacuum stability. Finally, we reserve \cref{sec:pheno} for a brief analysis of the phenomenological implications (focusing on \cp violation) and \cref{sec:conclusions} for our conclusions.  A Table listing the most relevant findings of our work may be found at the beginning of the Conclusions.

%% file: sec_2HDM.tex
As emphasized above, we focus on the scalar sector of the theory. In general, gauge invariance implies that the potential can only include bilinear and quartic terms. Each of the three bilinear terms has a corresponding mass parameter, of which two ($m_{11}^2$ and $m_{22}^2$) are real while the third, $m_{12}^2$, is associated with a bilinear mixing of both Higgs doublets  and may be complex. 

Regarding the quartic couplings in the scalar potential, the two associated with self interactions of each of the Higgs fields, $\lambda_1$ and $\lambda_2$, must be real and, due to vacuum stability, positive. There are two couplings associated with Hermitian combinations of the Higgs fields, $\lambda_3$ and $\lambda_4$, which must be real, though not necessarily positive.  The coupling $\lambda_5$ is associated with the square of the gauge invariant bilinear of both Higgs fields, and it may therefore be complex. The couplings $\lambda_6$ and $\lambda_7$ are associated with the product of Hermitian bilinears of each of the Higgs fields with the gauge invariant bilinear of the two Higgs fields, and, as with $\lambda_5$, they may be complex. The most general scalar potential for a complex 2HDM is therefore:
\begin{equation}\label{potential}
\begin{split}
	V & = m_{11}^2 \Phi_1^\dagger \Phi_1 + m_{22}^2 \Phi_2^\dagger \Phi_2 - (m_{12}^2 \Phi_1^\dagger \Phi_2 + h.c.)\\
	& + \frac{\lambda_1}{2} (\Phi_1^\dagger \Phi_1)^2 + \frac{\lambda_2}{2} (\Phi_2^\dagger \Phi_2)^2 + \lambda_3 (\Phi_1^\dagger \Phi_1)(\Phi_2^\dagger \Phi_2) + \lambda_4 (\Phi_1^\dagger \Phi_2) (\Phi_2^\dagger \Phi_1)\\
	& + \left[ \frac{\lambda_5}{2} (\Phi_1^\dagger \Phi_2)^2 + \lambda_6 (\Phi_1^\dagger \Phi_1)(\Phi_1^\dagger \Phi_2) + \lambda_7 (\Phi_2^\dagger \Phi_2) (\Phi_1^\dagger \Phi_2) + h.c. \right] \,,
\end{split}
\end{equation}
with $\Phi_{1,2} = (\Phi_{1,2}^+, \Phi_{1,2}^0)^T$ being complex $SU(2)$ doublets with hypercharge $+1$. 

One way to prevent Higgs-induced flavor violation in the fermion sector is to introduce a $\mathbb{Z}_2$ parity symmetry under which each charged fermion species transforms as even or odd. The Higgs doublets are assigned opposite parities and couple only to those charged fermions that carry their own parity. In such a scenario, the terms accompanying the couplings $\lambda_6$ and $\lambda_7$ would violate parity symmetry and hence should vanish. The mass parameter $m_{12}^2$ is also odd under the parity symmetry but induces only a soft breaking of this symmetry, which does not affect the ultraviolet properties of the theory. Thus it is usually allowed.

There are alternative ways of suppressing flavor violating couplings of the Higgs to fermions which do not rely on a simple parity symmetry and hence allow for the presence of $\lambda_6$ and $\lambda_7$ terms. One example is the flavor-aligned 2HDM~\cite{Pich:2009sp}. Alternatively, one can impose a parity symmetry in the ultraviolet but allow the effective low energy field theory to be affected by operators generated by the decoupling of a sector where this symmetry is broken softly by dimensionful couplings which do not respect the parity symmetry properties. One example of such a theory is the NMSSM in the presence of heavy singlets, as discussed in Ref.~\cite{Coyle:2018ydo}.  In this case, the presence of the couplings $\lambda_6$ and $\lambda_7$ is essential to allow for the alignment of the light Higgs boson with a SM-like Higgs, leading to a good agreement with precision Higgs physics even in the case of large Higgs self couplings.

So we see that it is not necessary to restrict to the $\mathbb{Z}_2$-symmetric 2HDM with vanishing $\lambda_6$ and $\lambda_7$ to avoid Higgs induced flavor violation in the fermion sector.\footnote{Note that non-vanishing $\lambda_{6,7}$ induces flavor violation via Higgs mixing. This effect is, however, loop suppressed.} Further, it is phenomenologically interesting to study the 2HDM in full generality with these terms present. One consequence would be the possibility of having charge-parity (\cp) violation in the bosonic sector. Indeed, to keep good agreement with Higgs precision data~\cite{CMS:2018uag,ATLAS:2019nkf},  one is normally interested in studying 2HDM in (or close to) the exact alignment limit --- the limit in which one of the neutral scalars carries the full vacuum expectation value and has SM-like tree-level couplings~\cite{Gunion:2002zf,Carena:2013ooa,BhupalDev:2014bir,Carena:2014nza,Carena:2015moc,Bernon:2015qea}. If one imposes exact alignment in the $\mathbb{Z}_2$-symmetric 2HDM, however, \cp is necessarily conserved, as will be explained in detail in \cref{sec:pheno}. In the full 2HDM, on the other hand, one can have \cp violation whilst remaining in exact alignment thanks to the presence of $\lambda_6$ and $\lambda_7$ terms. This \cp violation could manifest in the neutral scalar mass eigenstates as well as bosonic couplings (see also \ccite{Haber:2022gsn}), providing many potential experimental signatures. With this motivation in mind, we keep $\lambda_6$ and $\lambda_7$ non-zero throughout this work.

%% file: sec_math.tex
In this work, much of the analysis of perturbative unitarity and vacuum stability involves placing bounds on matrix eigenvalues. In the most general 2HDM, analytic expressions for these constraints are either very complicated or simply can not be formulated. To obtain some analytic insight, we derive conditions which are either necessary or sufficient. Their derivation is based on three linear algebra theorems which we briefly review here.


\subsection{Frobenius norm}\label{sec:math_frobenius}

One may derive a bound on the magnitude of the eigenvalues of a matrix using the matrix norm. The following definition and theorem are needed: \\

\begin{centering}
\textbf{Theorem:} \textit{The magnitude of the eigenvalues $e_i$ of a square matrix $A$ are bounded  from above by the matrix norm: $|e_i| \leq ||A||$.}
\end{centering} \\

\noindent where a matrix norm is defined as: \\

\noindent \textbf{Definition:} \textit{Given two $m \times n$ matrices $A$ and $B$, the matrix norm $||A||$ satisfies the following properties:}
\begin{itemize}
    \item $||A||\geq 0$,
    \item $||A|| = 0 \Leftrightarrow A=0_{m,n}$,
    \item $||\alpha A|| = |\alpha| ||A||$,
    \item $||A+B|| \leq ||A|| + ||B||$.
\end{itemize}

\noindent The above theorem holds for any choice of matrix norm, and thus one may employ the Frobenius norm ~\cite{Garren}, $||A|| = \sqrt{\text{Tr}(A^{\dagger}A)}$, to find the following result:
\begin{equation}
    |e_i| \leq \sqrt{\text{Tr}(A^{\dagger}A)}
\end{equation}

\noindent This bound on the eigenvalues will be used to derive sufficient bounds in the following sections. 

\subsection{Gershgorin disk theorem} \label{sufficienttheorem}

We will use the Gershgorin disk theorem~\cite{Gershgorin:1931} in upcoming sections to derive sufficient conditions for perturbative unitarity and vacuum stability of the 2HDM potential. The theorem is typically used to constrain the spectra of complex square matrices. The basic idea is that one identifies each of the diagonal elements with a point in the complex plane and then constructs a disk around this central point, with the radius given by the sum of the magnitudes of the other $n-1$ entries of the corresponding row.\footnote{One can also construct the radius by summing the magnitudes of the $n-1$ column entries.} The theorem says that all eigenvalues must lie within the union of these disks. Formally, we have the following definition and theorem:\\

\begin{centering}
\textbf{Definition:} \textit{Let $A$ be a complex $n\times n$ matrix with entries $a_{ij}$, and let $R_i$ be the sum of the magnitudes of the non-diagonal entries of the $i^{\text{th}}$ row, $R_i = \sum_{j \neq i} |a_{ij}|$. Then the Gershgorin disk $D(a_{ii}, R_i)$ is defined as the closed disk in the complex plane centered on $a_{ii}$ with radius $R_i$.}
\end{centering}\\

\begin{centering}
\textbf{Theorem:} \textit{Every eigenvalue of $A$ lies within at least one such Gershgorin disk $D(a_{ii}, R_i)$.}
\end{centering}

\noindent This theorem can be used to derive an upper bound on the magnitude of the eigenvalues of a matrix. We will use this technique below when discussing perturbative unitarity and vacuum stability. Since all the matrices we will consider in the subsequent sections on perturbative unitarity and boundedness from below are Hermitian matrices, each eigenvalue will lie within the intervals formed by the intersection of the Gershgorin disks with the real axis.

We shall proceed in the following manner: We will first construct the intervals containing the eigenvalues of each matrix $A$. For each interval, the rightmost and leftmost endpoints $x_i^\pm$ will be given by the sum and difference, respectively, of the center and the radius,
\begin{equation}
	x_{i}^\pm \equiv a_{ii} \pm R_i \,, \,\,\,\text{with}\,\,\, R_i = \sum_{j \neq i} |a_{ij}| \,.
\end{equation}
We then identify which $x_i^\pm$ extends furthest in the positive or negative direction. We know that every eigenvalue $e_k$ must lie within the endpoints of the largest possible total interval,
\begin{equation}
	\text{min}(x_i^-) \leq e_k \leq \text{max}(x_i^+) \,.
\end{equation}

\noindent This may be rephrased into an upper bound on $|e_k|$ as:
\begin{align}
    |e_k| \leq \mathrm{max}_i \left( \sum_{j} |a_{ij}| \right) ,
    \label{eq:ek}
\end{align}
where the left-hand side of \cref{eq:ek} represents the absolute value of any given eigenvalue $e_k$ and the right-hand side  represents the maximum value of $\sum_j |a_{ij}|$ over all rows $i$ of the matrix $A$. In fact, this condensed statement of Gershgorin circle theorem is an application of the matrix norm theorem, employing the norm $||A||=\text{max}_i(\sum_j |a_{ij}|)$.

\subsection{Principal minors} \label{necessarytheorem}

In order to derive necessary conditions, one may employ Sylvester's criterion in a clever way, as proposed in \ccite{Bento:2022vsb}. Sylvester's criterion involves the principal minors $D_k$ of a matrix, where $D_k$ is the determinant of the upper-left $k\times k$ sub-matrix. The statement of Sylvester's criterion is the following: \\

\begin{centering}
\textbf{Theorem:} \textit{Let M be a Hermitian $n \times n$ matrix. M is positive definite if and only if all of the principal minors $D_k(M)$ are positive.}
\end{centering} \\

\noindent We further need the following result about Hermitian matrices\footnote{This theorem is in fact used in the proof of Sylvester's criterion}: \\

\begin{centering}
\textbf{Theorem:} \textit{Let M be a Hermitian matrix. Then M is positive definite if and only if all of its eigenvalues are positive.}
\end{centering} \\

One can apply this to derive an upper bound on the eigenvalues of a diagonalizable matrix in the following way: \\

\begin{centering}
\textbf{Theorem:} \textit{Let M be an $n\times n$ diagonalizable, Hermitian (symmetric) matrix and let $c$ be a positive real number. The eigenvalues $e_i$ of M are bounded as $|e_i|<c$ if and only if all principal minors $D_k(c\mathbb{1} - M)$ and $D_k(c\mathbb{1} + M)$ are positive for all $k=1...n$.} \end{centering}\\

\noindent To see this, consider applying a unitary transformation which diagonalizes $M$ to the matrix $c\mathbb{1} \pm \mathrm{diag}(M)$. Then for symmetric or Hermitian matrices, the statement that $c\mathbb{1} \pm M$ is positive definite becomes a statement on the relative values of $e_i$ and c. In this manner, the application of Sylvester's criterion to these specific matrices allows one to put an upper bound on the magnitude of the eigenvalues without diagonalizing the matrix $M$. Note that for the absolute value $|e_i|$ to be bounded by $c$, we require the use of both $c\mathbb{1} \pm M$ matrices. On the other hand, if one has only an upper bound on $e_i$, i.e. $e_i<c$, as we will have in the case of vacuum stability, then one only requires the principal minors of the matrix $c\mathbb{1}-M$ to be positive. 

We note that the use specifically of the upper-left sub-matrices in Sylvester's criterion is an arbitrary choice, and basis-dependent. One could instead consider the lower-right sub-matrices, or any matrices along the diagonal. As such, it is possible to derive further conditions using this criteria by further considering, for example, the upper-left, lower-right, and central $2\times 2$ sub-matrices of a $4\times 4$ matrix. We will do so in later analyses to strengthen the lower-$k$ bounds.

This use of sub-determinants has been proposed in \ccite{Bento:2022vsb} as a method to increase the efficiency of parameter scans in models with large scattering matrices. For such theories (e.g.\ the model with $N$ Higgs doublets, NHDM, considered in \ccite{Bento:2022vsb} for higher $N$), the numerical calculation of the scattering matrix eigenvalues is computationally expensive. We note that the use of the Gershgorin disk theorem proposed in \cref{sufficienttheorem} provides an additional complementary method to speed-up parameter scans.

%% file: sec_unitarity.tex
Tree-level constraints for perturbative unitarity in the most general 2HDM have already been investigated in the literature~\cite{
Ginzburg:2005dt,Kanemura:2015ska,Jurciukonis:2018skr}. However, for a non-zero $\lambda_{6}$ and $\lambda_7$, no exact analytic conditions have been obtained yet. Here, we will first review the existing literature and then derive analytic expressions for the case of non-vanishing $\lambda_6$ and $\lambda_7$.


\subsection{Numerical bound}

Perturbative unitarity is usually imposed by demanding that the eigenvalues $e_i$ of the scalar scattering matrix at high energy be less than the unitarity limit, $|e_i| < 8 \pi$. Thus to derive the constraints on the quartic couplings, one must construct the scattering matrix for all physical scalar states.

We are interested in all processes $AB \rightarrow CD$, where the fields $A...D$ represent any combination of the physical\footnote{Technically since we are working in the high energy limit, the equivalence theorem allows us to replace $W_L^\pm$ and $Z_L$ by their corresponding Goldstone bosons.} scalars $(H_{1}, H_2, H_3, H^\pm, W_L^\pm, Z_L)$. The interactions and hence S-matrix take a complicated form in terms of the physical states. However since we are only interested in the eigenvalues of the S-matrix, we may choose any basis related to the physical basis by a unitary transformation. The derivation is simplest in the basis of the original Higgs fields $(w_i^\pm, h_i, z_i)$, appearing as
\begin{equation}\label{weak}
	\Phi_i = \begin{pmatrix} w_i^+ \\ \frac{1}{\sqrt{2}} (v_i + h_i + i z_i) \end{pmatrix} \,, 
\end{equation}
with $v = \sqrt{\sum_i v_i^2}$ = 246~GeV. \\
Out of these fields, we can construct 14 neutral two-body states: $\ket{w_i^+ w_i^-}$, $\frac{1}{\sqrt{2}}\ket{z_i z_i}$, $\frac{1}{\sqrt{2}}\ket{h_i h_i}$, $\ket{h_i z_i}$, $\ket{w_1^+ w_2^-}$, $\ket{w_2^+ w_1^-}$, $\ket{z_1 z_2}$, $\ket{h_1 h_2}$, $\ket{z_1 h_2}$, and $\ket{h_1 z_2}$. By constructing states which are linear combinations with definite hypercharge and total weak isospin, denoted by $(Y, I)$, and grouping the ones with the same set of quantum numbers, the matrix of S-wave amplitudes $a_0$ takes a block diagonal form (for more details see \ccite{Ginzburg:2005dt,Kanemura:2015ska,Jurciukonis:2018skr}). For the neutral scattering channels, this is:
\begin{equation}
	a_0^{(0)} = \frac{1}{16 \pi} \begin{pmatrix} X_{(0,0)} & & & \\ & X_{(0,1)} & & \\ & & X_{(1,1)} & \\ & & & X_{(1,1)} \end{pmatrix} \,,
\end{equation}
where the subscript of each submatrix denotes the quantum numbers $(Y, I)$ of the corresponding states. The entries are:
\begin{subequations}
\begin{align}
	X_{(0,0)} &= \begin{pmatrix} 3 \lambda_1 & 2 \lambda_3 + \lambda_4 & 3 \lambda_6 & 3 \lambda_6^* \\ 2 \lambda_3 + \lambda_4 & 3 \lambda_2 & 3 \lambda_7 & 3 \lambda_7^* \\ 3 \lambda_6^* & 3 \lambda_7^* & \lambda_3 + 2 \lambda_4 & 3 \lambda_5^* \\ 3 \lambda_6 & 3 \lambda_7 & 3 \lambda_5 & \lambda_3 + 2 \lambda_4 \end{pmatrix} \,, \\
	X_{(0,1)} &= \begin{pmatrix} \lambda_1 & \lambda_4 & \lambda_6 & \lambda_6^* \\ \lambda_4 & \lambda_2 & \lambda_7 & \lambda_7^* \\ \lambda_6^* & \lambda_7^* & \lambda_3 & \lambda_5^* \\ \lambda_6 & \lambda_7 & \lambda_5 & \lambda_3 \end{pmatrix} \,, \\
	X_{(1,1)} &= \begin{pmatrix} \lambda_1 & \lambda_5 & \sqrt{2}\lambda_6 \\ \lambda_5^* & \lambda_2 & \sqrt{2} \lambda_7^* \\ \sqrt{2} \lambda_6^* & \sqrt{2} \lambda_7 & \lambda_3 + \lambda_4 \end{pmatrix} \,.
\end{align}
\end{subequations}
For the 8 singly-charged two-body states $\ket{w_i^+ z_i}$, $\ket{w_i^+ h_i}$, $\ket{w_1^+ z_2}$, $\ket{w_1^+ h_2}$, $\ket{w_2^+ z_1}$, $\ket{w_2^+ h_1}$, the block diagonal 8$\times$8 singly-charged S-matrix is given by:
\begin{equation}
	a_0^{(+)} = \frac{1}{16 \pi} \begin{pmatrix} X_{(0,1)} & & \\ & X_{(1,0)} & \\ & & X_{(1,1)} \end{pmatrix} \,,
\end{equation}
where the new entry $X_{(1,0)}$ is just the one-dimensional eigenvalue:
\begin{equation}
	X_{(1,0)} = \lambda_3 - \lambda_4 \,.
\end{equation}
Finally, the 3$\times$3 S-matrix for the three doubly-charged 2-body states
	 $\ket{w_i^+ w_i^+}$, $\ket{w_1^+ w_2^+}$ is given by:
\begin{equation}
	a_0^{(++)} = \frac{1}{16 \pi} X_{(1,1)} \,.
\end{equation}
We impose perturbative unitarity by demanding that the eigenvalues of the scattering matrix are smaller than $8\pi$ implying that $|a_0| < \frac{1}{2}$. Indeed, the eigenvalues of the submatrices $X_{(0,0)}$, $X_{(0,1)}$, $X_{(1,0)}$, and $X_{(1,1)}$, which we denote as $e_i$, must all satisfy
\begin{equation} \label{eq:unitarity_exact}
	|e_i| < 8\pi \,.
\end{equation}
Obtaining analytic expressions for the eigenvalues requires solving cubic and quartic equations, and the result is complicated and not very useful. Given a choice of input parameters, however, it is easy to check this condition numerically.

Assuming all $\lambda_{1...7}$ to be equal, the strongest constraint arises from the 4$\times$4 matrix $X_{00}$. If we set all $\lambda_i \equiv \lambda$ and solve for the eigenvalues, we find the bound:
\begin{equation}
	\lambda < \frac{2\pi}{3} \,.
\end{equation}
This value is an order of magnitude smaller than $8 \pi$, implying that if all quartic couplings are sizable (i.e.\ of \order{1}), perturbative unitarity may be lost even at values of the couplings much smaller than $4 \pi$, which is a bound often encountered in the literature to ensure perturbativity.

\begin{figure}
    \centering
    \includegraphics[width=0.6\textwidth]{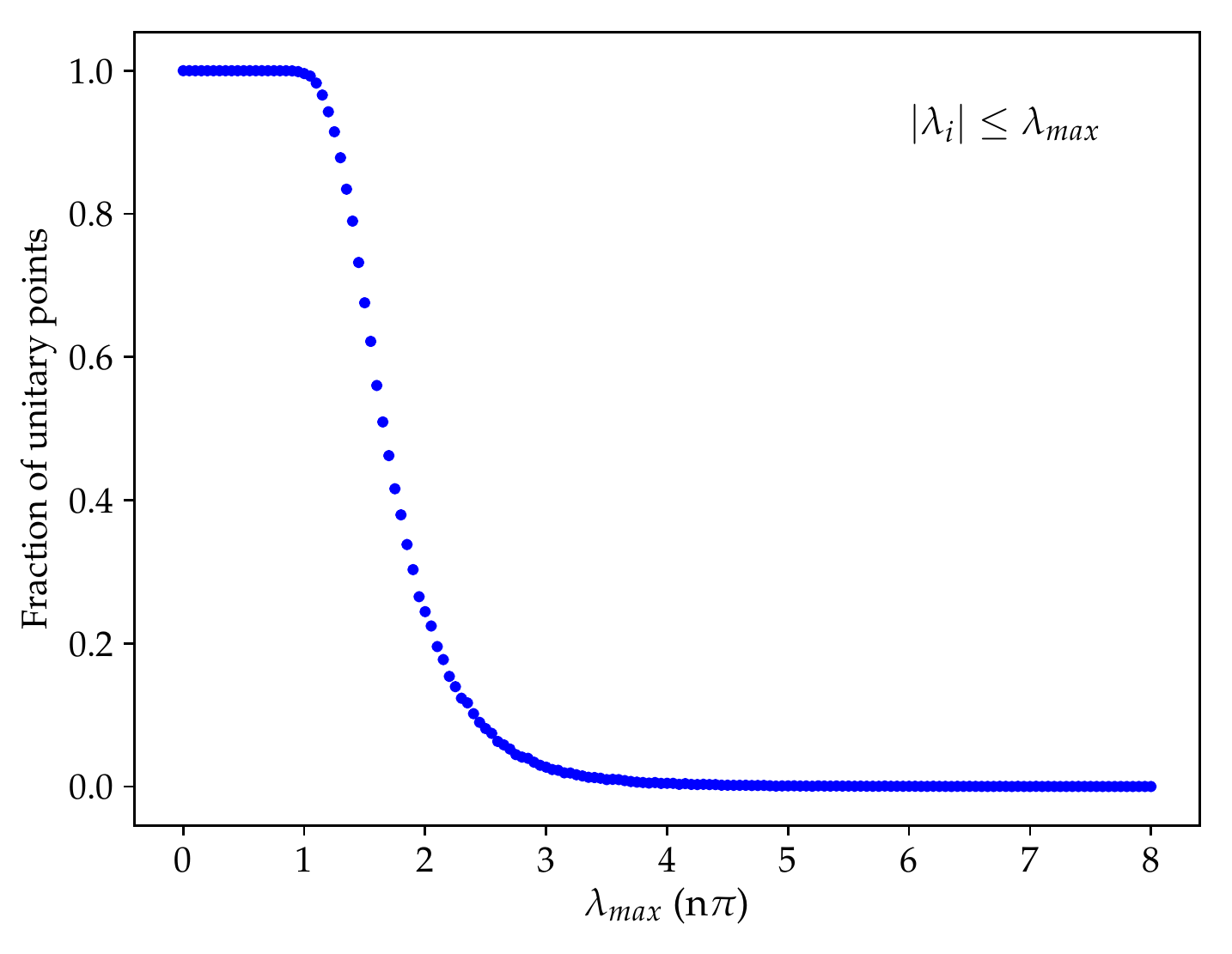}
    \caption{Plot showing the fraction of points that pass the unitarity bound $|e_i|<8\pi$ for different choices of $\lambda_{max}$, in units of multiples of $\pi$. The values of the $\lambda_i$ are each chosen randomly such that $|\lambda_i|<\lambda_{max}$. We test 20,000 random sets of $\lambda_i$ for each $\lambda_{max}$.} 
    \label{fig:lambda_unitary}
\end{figure}

We investigate the validity of such an upper bound further in \cref{fig:lambda_unitary}. For this figure, we randomly choose each $\lambda_i$ within the range $|\lambda_i|<\lambda_\text{max}$ and then show the fraction of test points which pass the numerical unitarity constraint, as a function of $\lambda_\text{max}$. For $\lambda_\text{max} \lesssim \pi$ almost all points survive the perturbative unitarity constraint. For larger $\lambda_\text{max}$ values the survival rate quickly drops to almost zero for $\lambda_\text{max} \gtrsim 4\pi$. This highlights again that the simplified perturbativity bound of $|\lambda_i| < 4\pi$, which is often encountered in the literature, is too loose. Based on the results in \cref{fig:lambda_unitary}, a better choice of bound might be $|\lambda_i|\lesssim \pi$ or $|\lambda_i|\lesssim 3\pi/2$.


\subsection{A necessary condition for perturbative unitarity}

To gain some intuition for the perturbative unitarity constraint, we now turn to derive some simplified analytic conditions which are either necessary or sufficient, though not both. In this section we will focus on the former, which can be used to quickly rule out invalid parameter sets which violate perturbative unitarity. One can derive necessary conditions by invoking the method of principal minors, which is reviewed in \cref{necessarytheorem} and can be used to give an upper bound on the maximal value of the eigenvalues of a Hermitian matrix. Since the scattering matrices are Hermitian, demanding the eigenvalues to be bounded as $|e_i|<8\pi$, as required by perturbative unitarity, amounts to requiring:
\begin{equation}
   D_k(8\pi \mathbb{1} + X) > 0 \,\, \text{and} \,\, D_k(8\pi \mathbb{1} - X) > 0\,,
\end{equation}
for $k=1,2,3,4$. Since satisfying both criteria for all $k=1,2,3,4$ is a necessary and sufficient condition, any single $k$ condition provides a necessary condition.

Since the eigenvalues of $X_{(0,0)}$ are generically the largest and therefore the most constraining, we will focus on bounds coming from this matrix. We begin with the upper left $2\times 2$ submatrices. Taking the determinant, we have:
\begin{subequations}\label{eq:unitarity_D2}
\begin{equation} 
    D_{2}^L(8\pi \mathbb{1} + X_{(0,0)})>0\,\,\, \Rightarrow\,\,\, 64 \pi^2 + 24\pi (\lambda_1 + \lambda_2) + 9\lambda_1 \lambda_2 - (2 \lambda_3 + \lambda_4)^2 > 0 \,,
\end{equation}
\begin{equation}\label{eq:D2UL}
    D_{2}^L(8\pi \mathbb{1} - X_{(0,0)}) > 0 \,\,\, \Rightarrow \,\,\, 64 \pi^2 - 24\pi (\lambda_1 + \lambda_2) + 9\lambda_1 \lambda_2 - (2 \lambda_3 + \lambda_4)^2 > 0 \,.
\end{equation}
\end{subequations}
Clearly the latter constraint coming from $D_{2}^L(8\pi \mathbb{1}-X)$ will be the stronger of the two, since $\lambda_1, \lambda_2 > 0$ if boundedness from below is imposed. Thus the necessary $k=2$ condition reduces to~\cref{eq:D2UL}. We also examine the constraints that arise from using the lower-right and center $2\times2$ sub-matrices, as proposed in \cref{necessarytheorem}. The analytic conditions from the lower-right $D^R$ and center $D^C$ sub-matrices are, respectively,
\begin{subequations}
\begin{align}
    D_{2}^{R}(8\pi\mathbb{1} - X_{(0,0)}) &= 64\pi^2 + (\lambda_3 - 16\pi)\lambda_3+ (4(\lambda_3 + \lambda_4) - 32\pi)\lambda_4 - 9|\lambda_5|^2 &> 0\,, \\
    D_{2}^{C}(8\pi\mathbb{1} - X_{(0,0)}) &= 64\pi^2  - 8\pi(\lambda_3 + 3\lambda_2 + 2\lambda_4) + 3\lambda_3\lambda_2 + 6\lambda_2\lambda_4  - 9|\lambda_7|^2 &> 0\,.
\end{align}
\end{subequations}
The combination of these three expressions provides a stronger constraint than just the upper-left minor constraint alone.

While it is immediately clear that $D_{2}^L(8\pi \mathbb{1}-X)$ is stronger than the addition-based bound for the upper-left matrix, it is not clear for the center and lower-right matrices; in fact, including these bounds provides a slightly more constraining result. We thus employ the $D_{2}^{C,R}(8\pi \mathbb{1}+X)$ constraints in our analysis of the $k=2$ bound as well:
\begin{subequations}
\begin{align}
    D_{2}^{R}(8\pi\mathbb{1} + X_{(0,0)}) &= 64\pi^2 + (\lambda_3 + 16\pi)\lambda_3+ (4(\lambda_3 + \lambda_4) + 32\pi)\lambda_4 - 9|\lambda_5|^2 &> 0\,, \\
    D_{2}^{C}(8\pi\mathbb{1} + X_{(0,0)}) &= 64\pi^2  + 8\pi(\lambda_3 + 3\lambda_2 + 2\lambda_4) + 3\lambda_3\lambda_2 + 6\lambda_2\lambda_4  - 9|\lambda_7|^2 &> 0\,.
\end{align}
\end{subequations}

\begin{figure}
    \centering
    \includegraphics[width=0.6\textwidth]{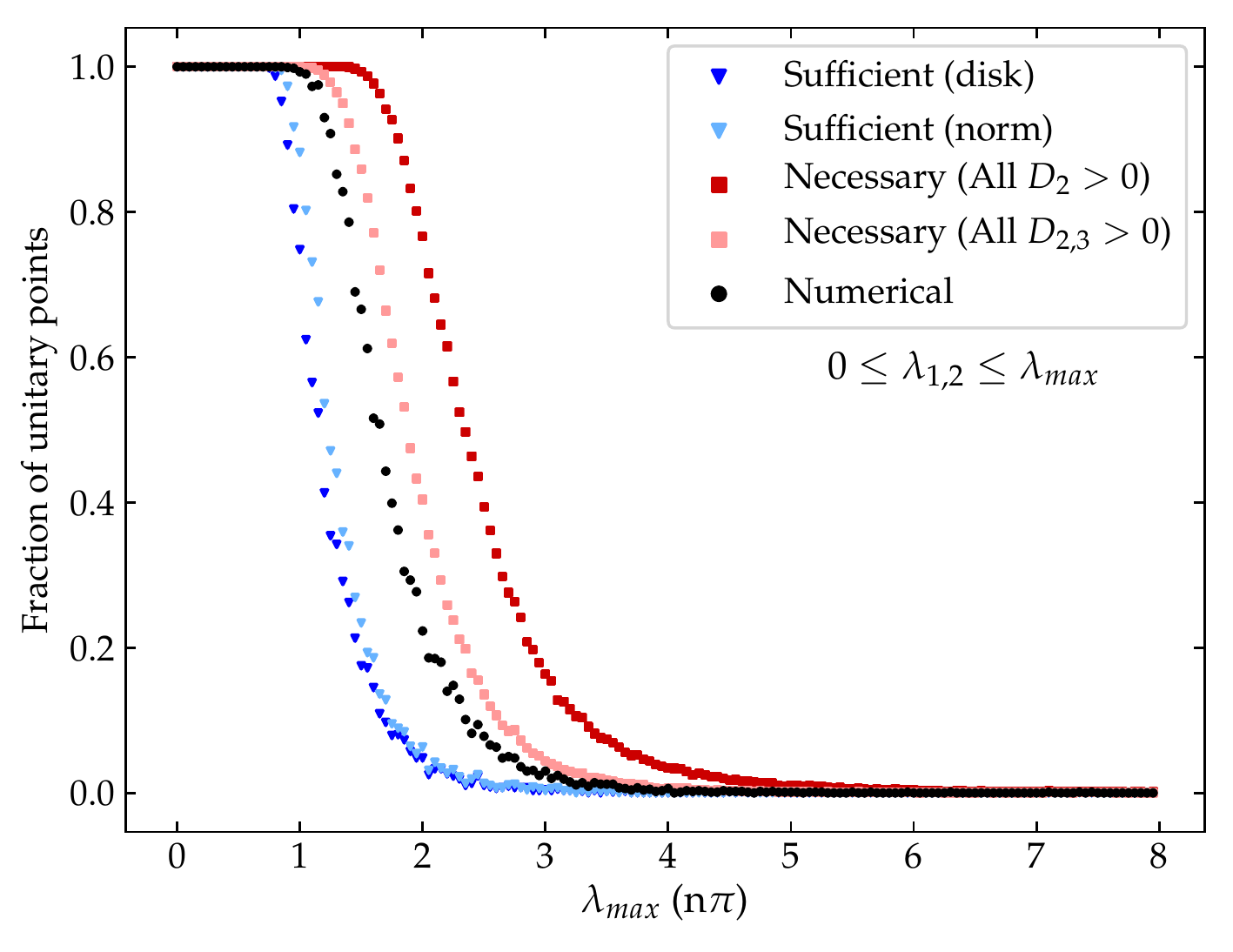}
    \caption{Plot comparing the number of points that pass the exact numerical bound $|e_i|<8\pi$ (black), the sufficient bound from the Gershgorin disk theorem \cref{unitarityGerschgorin} (dark blue), the sufficient bound from the Frobenius norm \cref{eq:unitarityFrobenius} (light blue), the necessary condition $D_2(8\pi\mathbb{1}\pm X_{(0,0)})>0$ (dark red), and the necessary condition $D_{2,3}(8\pi\mathbb{1}\pm X_{(0,0)})>0$ (light red). The $\lambda_{i}$ values are randomly chosen from the range of values satisfying $|\lambda_{i}|<\lambda_{max}$, where $\lambda_{max}$ is given by the x-axis in units of multiples of $\pi$. The minimal bounded from below condition $\lambda_{1,2}\geq0$ is enforced. The $\lambda_{5,6,7}$ values are allowed to be complex. The total number of points checked for each $\lambda_{max}$ is 10,000. }
    \label{fig:unitary_conditions}
\end{figure}

Next, we look at the upper left $3\times 3$ submatrices. Unlike the $k=2$ case, it is not clear that one of these is generically more constraining than the other. To be consistent with the $k=2$ case, we will examine the $8\pi\mathbb{1}-X_{(0,0)}$ matrix. We additionally consider the lower-right $3\times 3$ sub-matrix. These give the following bounds:
\begin{subequations}\label{eq:unitarity_D3}
\begin{equation}\label{eq:unitarity_D3_1}
\begin{split}
    D_3^L(8 \pi \mathbb{1}-X_{(0,0)}) =& (8\pi-\lambda_3-2\lambda_4)((8\pi-3\lambda_1)(8\pi-3\lambda_2)-(2\lambda_3+\lambda_4)^2) \\
    &-9(8\pi-3\lambda_2)|\lambda_6|^2 -9(8\pi-3\lambda_1)|\lambda_7|^2\\
    &-9(2\lambda_3+\lambda_4)(\lambda_6\lambda_7^*+\lambda_6^*\lambda_7 ) > 0 \,,
\end{split}
\end{equation}
\begin{equation}\label{eq:unitarity_D3_2}
\begin{split}
    D_3^{R}(8\pi\mathbb{1}-X_{(0,0)}) =& (8\pi - \lambda_3-2\lambda_4 )^2(8\pi-3\lambda_2) - 9|\lambda_5|^2 (8\pi-3\lambda_2) \\
    &- 27 (\lambda_5^* \lambda_7^2 + \lambda_5 (\lambda_7^*)^2) - 18(8\pi - \lambda_3-2\lambda_4)|\lambda_7|^2 > 0\,.
\end{split}
\end{equation}
\end{subequations}
Meanwhile, considering $8\pi \mathbb{1} + X_{(0,0)}$ gives
\begin{subequations}\label{eq:unitarity_D3_plis}
\begin{equation}\label{eq:unitarity_D3_plus_1}
\begin{split}
    D_3^L(8 \pi \mathbb{1}+X_{(0,0)}) =& (8\pi+\lambda_3+2\lambda_4)((8\pi+3\lambda_1)(8\pi+3\lambda_2)-(2\lambda_3+\lambda_4)^2) \\
    &-9(8\pi+3\lambda_2)|\lambda_6|^2 -9(8\pi+3\lambda_1)|\lambda_7|^2\\
    &+9(2\lambda_3+\lambda_4)(\lambda_6\lambda_7^*+\lambda_6^*\lambda_7 ) > 0 \,,
\end{split}
\end{equation}
\begin{equation}\label{eq:unitarity_D3_plus_2}
\begin{split}
    D_3^{R}(8\pi\mathbb{1}+X_{(0,0)}) =& (8\pi + \lambda_3 + 2\lambda_4 )^2(8\pi+3\lambda_2) - 9|\lambda_5|^2 (8\pi+3\lambda_2) \\
    &+ 27 (\lambda_5^* \lambda_7^2 + \lambda_5 (\lambda_7^*)^2) - 18(8\pi + \lambda_3+2\lambda_4)|\lambda_7|^2 > 0\,.
\end{split}
\end{equation}
\end{subequations}
While the $D_3(8\pi \mathbb{1} - X_{(0,0)})$ provide the strongest constraints for values of $|\lambda_i|\lesssim 4\pi$, the inclusion of the $D_3(8\pi \mathbb{1} + X_{(0,0)})$ and $D_2$ bounds improves the performance of the $k=3$ bounds at higher $|\lambda_i|$.
We omit analytic expressions for the $k=4$ case, since they cannot be simplified to a useful form. Moreover, the $k=3$ expressions already provide constraints very close to the full numerical bound (see \cref{fig:unitary_conditions}).


\subsection{Sufficient conditions for perturbative unitarity}
\label{sec:unitarity_sufficent}

Next, we turn to derive sufficient conditions for perturbative unitarity by applying the Gershgorin disk theorem, which is reviewed in \cref{sufficienttheorem} and gives an upper bound on the maximal value of the eigenvalues. By demanding that this upper bound is less than $8\pi$, we obtain a sufficient condition for perturbative unitarity.

We first construct the intervals $x_i^{(Y,I)}$ containing the eigenvalues of each of the scattering matrices, $X_{(0,0)}$, $X_{(0,1)}$, $X_{(1,0)}$, and $X_{(1,1)}$. We know that in order to uphold perturbative unitarity, we must have $|e_i| < 8\pi$. Thus we arrive at the sufficient condition:
\begin{equation}
	\text{max}( x_i^{(Y,I)} ) < 8\pi \,.
\end{equation}
For each of the $X$ matrices, we can work out the $x_i^{(Y,I)}$ explicitly. For $X_{(0,0)}$, we obtain:
\begin{subequations}
\begin{align}
	x_{1}^{(0,0)} &= 3 |\lambda_1| + (|2 \lambda_3 + \lambda_4| + 6 |\lambda_6|) \,, \\
	x_{2}^{(0,0)} &= 3 |\lambda_2| + (|2 \lambda_3 + \lambda_4| + 6 |\lambda_7|) \,, \\
	x_{3}^{(0,0)} &= x_{4}^{(0,0)} = |\lambda_3 + 2 \lambda_4| + 3 ( |\lambda_5| + |\lambda_6| + |\lambda_7|) \,.
\end{align}
\end{subequations}
For $X_{(0,1)}$, they are:
\begin{subequations}
\begin{align}
	x_{1}^{(0,1)} &= |\lambda_1| + (|\lambda_4| + 2 |\lambda_6|) \,, \\
	x_{2}^{(0,1)} &= |\lambda_2| + (|\lambda_4| + 2 |\lambda_7|) \,, \\
	x_{3}^{(0,1)} &= x_{4}^{(0,1)} = |\lambda_3| + (|\lambda_5| + |\lambda_6| + |\lambda_7|) \,.
\end{align}
\end{subequations}
For $X_{(1,1)}$, we have:
\begin{subequations}
\begin{align}
	x_{1}^{(1,1)} &= |\lambda_1| + (|\lambda_5| + \sqrt{2} |\lambda_6|) \,, \\
	x_{2}^{(1,1)} &= |\lambda_2| + (|\lambda_5| + \sqrt{2} |\lambda_7|) \,, \\
	x_{3}^{(1,1)} &= |\lambda_3 + \lambda_4| + \sqrt{2} (|\lambda_6| + |\lambda_7|) \,.
\end{align}
\end{subequations}
Finally, $X_{(1,0)}$, we have:
\begin{align}
	x_{1}^{(1,0)} &= |\lambda_3 - \lambda_4| \,.
\end{align}
In examining these conditions, the leading coefficient of 3 in the first set suggests that the $x_i^{(0,0)}$ corresponding to $X_{(0,0)}$ will generically be larger than those corresponding to $X_{(0,1)}$, $X_{(1,0)}$, and $X_{(1,1)}$; a numerical check confirms this intuition. Thus the sufficient condition for perturbative unitarity simplifies slightly to 
\begin{equation}\label{unitarityGerschgorin}
	\text{max}\big(x_{i}^{(0,0)} \big) < 8\pi \,.
\end{equation} 

One may alternatively employ the bound arising from the Frobenius norm, as discussed in \cref{sec:math_frobenius}. Taking the dominant $X_{(0,0)}$ matrix, one finds the condition:
\begin{align}\label{eq:unitarityFrobenius}
    \sqrt{9(\lambda_1^2+\lambda_2^2) + 10(\lambda_3^2 + \lambda_4^2) + 16\lambda_3\lambda_4 + 18(|\lambda_5|^2 + 2|\lambda_6|^2 + 2|\lambda_7|^2)} \leq 8\pi
\end{align}
The dependence here on the signs of the $\lambda_i$ is similar to the dependence seen in the Gershgorin disk conditions: the bound is not sensitive to the signs of any $\lambda_i$ except for the relative sign between $\lambda_{3}$ and $\lambda_4$.


\subsection{Numerical comparison}


In order to compare the various bounds derived in this section, in \cref{fig:unitary_conditions} we plot the number of points which pass the exact, sufficient, and necessary conditions for different values of $\lambda_{max}$. For each $\lambda_{max}$, we consider 10,000 randomly-drawn values for the $\lambda_i$ within the range $|\lambda_i|<\lambda_{max}$. For the necessary conditions, the results are derived from the combination of all possible $2\times2$ ($3\times3$) sub-matrices along the diagonal for the $D_{2(3)}$ bound. We enforce the minimal bounded from below condition $\lambda_{1,2}>0$, which has been derived e.g.\ in \ccite{Branco:2011iw}. We find that the necessary $D_3(8\pi\mathbb{1}-X_{(0,0)})$ condition lies very close to the exact condition and is effective at ruling out parameter sets which fail perturbative unitarity, while for $\lambda_{max}\lesssim \pi$ all tested points satisfy perturbative unitarity.

%% file: sec_BFB.tex
Next, we seek to determine the conditions on the parameters such that the potential of \cref{potential} is bounded from below (BFB).  For this, it is necessary to ensure that the quartic part of the potential does not acquire negative values. If negative values were present, one could easily find indefinite negative values of the potential by rescaling all fields to infinity in the same direction as the one in which the negative value was found. We remark that analytic expressions have been formulated previously in \ccite{Song:2022gsz}, though for the case of explicit \cp conservation. We will make no such assumption. There are also previous analyses of the BFB condition using the eigenvalues of a 4$\times$4 matrix (\ccite{Ivanov:2006yq}); however, these analyses do not lead to analytical expressions, and we will follow an alternative approach.

We begin by reparameterizing the potential via $\Phi_1^\dagger \Phi_1 = \frac{1}{2} h_1^2$, $\Phi_2^\dagger \Phi_2 = \frac{1}{2} h_2^2$, $\Phi_1^\dagger \Phi_2 = \frac{1}{2} h_1 h_2 \rho e^{i \eta}$, with $\rho \in [0,1]$. Moreover, we decompose the complex couplings $\lambda_5$, $\lambda_6$, $\lambda_7$ into real and imaginary parts as: $\lambda_i e^{i\eta} + \lambda_i^* e^{-i \eta} = 2 \text{Re}[\lambda_i] \cos \eta - 2 \text{Im}[\lambda_i] \sin \eta$. The quartic part of the potential then becomes:
\begin{equation}\label{eq:Vquartic}
\begin{split}
	 V_{\text{quartic}} ={}\frac{1}{4}\bigg\{& \frac{\lambda_1}{2} h_1^4 + \frac{\lambda_2}{2} h_2^4 + \big[ \lambda_3 + \big( \lambda_4 + \text{Re}[\lambda_5]\cos 2 \eta - \text{Im}[\lambda_5] \sin 2 \eta \big) \rho^2 \big] h_1^2 h_2^2\\
	& + 2 \big( \text{Re}[\lambda_6] \cos \eta - \text{Im}[\lambda_6] \sin \eta \big) \rho\, h_1^3 h_2 \\
	& + 2 \big( \text{Re}[\lambda_7] \cos \eta - \text{Im}[\lambda_7] \sin \eta \big) \rho\, h_1 h_2^3 \bigg\}\,.
\end{split}
\end{equation}
We can then cast $V_{\text{quartic}}$ into the form
\begin{equation}
    V_{\text{quartic}} =
	\frac{1}{4}h_2^4 \left[ a \left( \frac{h_1}{h_2} \right)^4 + b \left(\frac{h_1}{h_2} \right)^3 + c \left(\frac{h_1}{h_2} \right)^2 + d \left(\frac{h_1}{h_2} \right) + e \right] \,,
\end{equation}
with 
\begin{subequations}\label{BFB_definitions1}
\begin{align}
	a &= \frac{\lambda_1}{2} , \hspace{.3cm} \ e = \frac{\lambda_2}{2} \, , \\
	b &= 2     \big( \text{Re}[\lambda_6] \cos \eta - \text{Im}[\lambda_6] \sin \eta \big) \rho \,, \\
	c &= \left[ \lambda_3 + \big( \lambda_4 + \text{Re}[\lambda_5]\cos 2 \eta - \text{Im}[\lambda_5] \sin 2 \eta \big) \rho^2 \right] \,, \\
	d &= 2 \big( \text{Re}[\lambda_7] \cos \eta - \text{Im}[\lambda_7] \sin \eta \big) \rho \, .
\end{align}
\end{subequations}
Clearly $a>0$ and $e>0$ has to be fulfilled as a minimum condition for BFB. We can then divide out by $e\, h_2^4$ and define the simplified polynomial:
\begin{equation} \label{eq:Vquartic_x}
    f(x) = x^4 + \alpha x^3 + \beta x^2 + \gamma x + 1 \,,
\end{equation}
with $x = \frac{a^{1/4}}{e^{1/4}} \frac{h_1}{h_2}$ and:
\begin{equation}
    \alpha = b a^{-3/4} e^{-1/4},\hspace{.5cm} \beta = c a^{-1/2} e^{-1/2},\hspace{.5cm} \gamma = d a^{-1/4} e^{-3/4} \,.
\end{equation}
We then define the following quantities:
\begin{subequations}\label{BFB_definitions2}
\begin{align}
	\Delta &= 4 [\beta^2 - 3 \alpha \gamma + 12]^3 - [72 \beta + 9 \alpha \beta \gamma - 2 \beta^3 -27 \alpha^2 - 27 \gamma^2]^2 \,,\\
	\chi_1 &= (\alpha - \gamma)^2 - 16 (\alpha + \beta + \gamma + 2) \,, \,\,\, \\
	\chi_2 &= (\alpha - \gamma)^2 - \frac{4(\beta+2)}{\sqrt{\beta - 2}} \left( \alpha + \gamma + 4 \sqrt{\beta - 2} \right) \,.
\end{align}
\end{subequations}
The positivity of $V_\text{quartic}$ is ensured if and only if one of the following conditions holds~\cite{doi:10.1137/0915035}:
\begin{equation}\label{positivityog}
\begin{split}
	&(1) \,\,\, \beta < - 2 \,\, \text{and} \,\, \Delta \leq 0 \,\, \text{and} \,\, \alpha + \gamma > 0 \,,\\
	&(2) \,\,\, -2 \leq \beta \leq 6 \,\, \text{and} \,\, \Delta \leq 0 \,\, \text{and} \,\, \alpha + \gamma > 0 \,, \\
	&(3) \,\,\, -2 \leq \beta \leq 6 \,\, \text{and} \,\, \Delta \geq 0 \,\, \text{and} \,\, \chi_1 \leq 0 \,, \\
	&(4) \,\,\, \beta > 6 \,\, \text{and} \,\, \Delta \leq 0 \,\, \text{and} \,\, \alpha + \gamma > 0 \,, \\
	&(5) \,\,\, \beta > 6 \,\, \text{and} \,\, \alpha >0 \,\, \text{and} \,\, \gamma >0 \,, \\
	&(6) \,\,\, \beta > 6 \,\, \text{and} \,\, \Delta \geq 0 \,\, \text{and} \,\, \chi_2 \leq 0 \,.
\end{split}
\end{equation}
If any of these conditions is true for a given set of input parameters $\lambda_{1...7}$ and for all possible values of $\rho \in [0,1]$, $\eta \in [0,2\pi)$, then the potential is BFB.

Note that under the transformation $\eta \rightarrow \eta + \pi$, both $\alpha$ and $\gamma$ are anti-symmetric ($\alpha \rightarrow - \alpha$ and $\gamma \rightarrow - \gamma$). This in turn implies that conditions (1), (2), (4), and (5) are always violated for some value of $\eta$, and therefore can never guarantee the positivity of $V_\text{quartic}$. Consequently, we are left with only two conditions under which the potential is BFB:
\begin{equation}\label{necessaryandsufficient}
	\text{The potential is BFB if and only if}: \,\, \Delta \geq 0 \,\, \text{and} \,\, \begin{cases} -2 \leq \beta \leq 6 \,\, \text{and} \,\, \chi_1 \leq 0 \,, \text{or}\\ \beta > 6 \,\, \text{and} \,\, \chi_2 \leq 0 \,. \end{cases}
\end{equation}
Note that upon setting $\lambda_6 = \lambda_7 = 0$, and after extremizing with respect to $\eta$, \cref{necessaryandsufficient} becomes
\begin{equation}
\begin{split}
\beta + 2 \geq 0 \; \Rightarrow	\; \lambda_3 + \rho^2 (\lambda_4 - |\lambda_5|) \geq - \sqrt{\lambda_1 \lambda_2} \, .
\end{split}
\end{equation}
This is a monotonic function of $\rho$, and hence the strongest constraints are derived for either $\rho=1$ or $\rho=0$, namely
\begin{align}
\lambda_3 +  \lambda_4 - |\lambda_5| & \geq - \sqrt{\lambda_1 \lambda_2} \,, \;\; \textrm{and}
\\
\lambda_3  & \geq - \sqrt{\lambda_1 \lambda_2} \,,
\label{eq:rhoeq0}
\end{align}
These conditions reproduce the well-known conditions for BFB in the $\mathbb{Z}_2$-symmetric 2HDM~\cite{Branco:2011iw}. Let us also stress that, for $\rho = 0$, \cref{necessaryandsufficient} leads to \cref{eq:rhoeq0} independently of the value of the other quartic couplings and hence this equation is a necessary condition for the potential stability even in the generic 2HDM case.


\subsection{Necessary conditions for boundedness from below} \label{sec:BFB_necessary}

The two options of \cref{necessaryandsufficient} present a necessary and sufficient condition for BFB. In order to implement this bound, one should scan over all possible values of $\rho$ and $\eta$, which can be computationally expensive for large parameter spaces. Thus we present here two simplified necessary (though not sufficient) conditions which can be used to quickly rule out invalid parameter sets and speed up scans.

We can first derive generalized versions of the existing literature bounds \cite{Branco:2011iw} by setting $x=1$ and taking $\rho = 1$ and $\eta=\frac{n\pi}{4}$, with $n=\{0,...,7\}$, in \cref{eq:Vquartic} and \cref{eq:Vquartic_x}. Applying this procedure to \cref{eq:Vquartic} leads to the following conditions:
\begin{subequations} \label{eq:BFBnec2}
    \begin{align}
        &\frac{\lambda_1 + \lambda_2}{2} + \lambda_3 + \lambda_4 + \lambda_5^R - 2|\lambda_6^R + \lambda_7^R| > 0 \,, \\
        &\frac{\lambda_1 + \lambda_2}{2} + \lambda_3 + \lambda_4 - \lambda_5^R - 2|\lambda_6^I + \lambda_7^I| > 0 \,, \\
        &\frac{\lambda_1 + \lambda_2}{2} + \lambda_3 + \lambda_4 + \lambda_5^I - \sqrt{2}\; \Big| (\lambda_6^R + \lambda_7^R) + (\lambda_6^I + \lambda_7^I)\Big| > 0 \,, \\
        &\frac{\lambda_1 + \lambda_2}{2} + \lambda_3 + \lambda_4 - \lambda_5^I - \sqrt{2}\; \Big| (\lambda_6^R + \lambda_7^R) - (\lambda_6^I + \lambda_7^I)\Big| > 0 \,,
    \end{align}
\end{subequations}
while applying  the same procedure to \cref{eq:Vquartic_x} leads to the conditions:
\begin{subequations}\label{eq:BFBnec3}
\begin{align}
    &\sqrt{\lambda_1\lambda_2}+\lambda_3 + \lambda_4 + \lambda_5^R - 2\left| \tilde{\lambda}_6^R +  \tilde{\lambda}_7^R \right| > 0 \,, \\
    &\sqrt{\lambda_1\lambda_2}+\lambda_3 + \lambda_4 - \lambda_5^R - 2\left| \tilde{\lambda}_6^I +  \tilde{\lambda}_7^I \right| > 0 \,, \\
    &\sqrt{\lambda_1\lambda_2}+\lambda_3 + \lambda_4 + \lambda_5^I - \sqrt{2}\left| (\tilde{\lambda}_6^R +  \tilde{\lambda}_7^R ) + (\tilde{\lambda}_6^I +  \tilde{\lambda}_7^I ) \right| > 0 \,, \\
    &\sqrt{\lambda_1\lambda_2}+\lambda_3 + \lambda_4 - \lambda_5^I - \sqrt{2}\left| (\tilde{\lambda}_6^R +  \tilde{\lambda}_7^R ) - (\tilde{\lambda}_6^I +  \tilde{\lambda}_7^I ) \right| > 0 \,.
\end{align}
\end{subequations}
Note that we have combined the $\eta,\eta+\pi$ conditions in each set to obtain four conditions instead of eight.

Alternatively, we can collapse the two conditions of \cref{necessaryandsufficient} into a single necessary condition as follows. Consider the two different branches with $\chi_{1,2} < 0$. Under the transformation $\eta \rightarrow \eta + \pi$, $\chi_1 \leq 0$ produces two conditions that must be satisfied simultaneously: $(\alpha - \gamma)^2 - 16(\alpha + \beta + \gamma + 2) \leq 0$ and $(\alpha - \gamma)^2 - 16( - \alpha + \beta - \gamma + 2) \leq 0$. We can add these together to obtain the simplified condition: $(\alpha - \gamma)^2 \leq 16(\beta+2)$. Similarly, demanding that $\chi_2 \leq 0$ for both $\eta$ and $\eta+ \pi$ gives us the simplified condition $(\alpha - \gamma)^2 \leq 16(\beta+2)$. So, we see that demanding $\chi_1 \leq 0$ and $\chi_2 \leq 0$ are equivalent, and both translate to the constraint:
\begin{equation}
	\chi_1 \leq 0 \,, \chi_2 \leq 0 \,\, \Rightarrow \,\, (\alpha - \gamma)^2 \leq 16(\beta+2) \,.
\end{equation}
In this way, the condition for the potential to be BFB can be reduced to the form:
\begin{equation}
	\Delta \geq 0 \,\, \text{and} \,\, \beta \geq -2 \,\, \text{and} \,\, (\alpha - \gamma)^2 \leq 16(\beta+2) \,.
\end{equation}
Note that both $\beta \geq -2$ and $(\alpha - \gamma)^2 \leq 16(\beta+2)$ restrict $\beta$, but that the latter will always be a stronger condition since $(\alpha-\gamma)^2 \geq 0$. Then this necessary but not sufficient BFB condition simplifies further to
\begin{equation}\label{necessary}
	\Delta \geq 0 \,\, \text{and} \,\, \beta \geq \frac{1}{16}(\alpha - \gamma)^2 -2 \,.
\end{equation}

This condition still depends on $\rho$ and $\eta$. Without loss of generality, we set\footnote{This gives us one necessary condition. We could obtain others by choosing $\rho < 1$, but these tend to be less constraining in most cases.} $\rho=1$. As for $\eta$, we need to find the value which extremizes the expression for each condition. Take for instance the latter condition of \cref{necessary} and define
\begin{equation}
	f(\eta) \equiv \beta - \frac{1}{16}(\alpha - \gamma)^2 + 2 \geq 0 \,.
\end{equation}
Using the definitions of \cref{BFB_definitions1,BFB_definitions2}, we can recast everything in terms of $\cos 2\eta$ and $\sin 2 \eta$ such that $f(\eta)$ only depends on these quantities. We can then easily determine the extremal value of $\eta_{\text{min}}$ which gives the minimal $f_\text{min}$. After some algebra, the positivity condition $f_{\text{min}} \geq 0$ reads:
\begin{equation}\label{BFB_necessary}
	2 (\lambda_1 \lambda_2 + \sqrt{\lambda_1 \lambda_2} (\lambda_3 + \lambda_4)) - \frac{1}{2} \big|\tilde{\lambda}_6 - \tilde{\lambda}_7 \big|^2 - \big| 2 \sqrt{\lambda_1 \lambda_2} \, \lambda_5 - \frac{1}{2} (\tilde{\lambda}_6 - \tilde{\lambda}_7)^2 \big| \geq 0 \,,
\end{equation}
where we have defined the rescaled couplings:
\begin{equation}
	\tilde{\lambda}_6 \equiv \left( \frac{\lambda_2}{\lambda_1} \right)^{1/4} \lambda_6 \,, \,\,\, \tilde{\lambda}_7 \equiv \left( \frac{\lambda_1}{\lambda_2} \right)^{1/4} \lambda_7 \,.
\end{equation}
Eqs.~(\ref{eq:BFBnec2}), (\ref{eq:BFBnec3}), and (\ref{BFB_necessary}) present simplified necessary conditions for BFB and are the main result of this section.


\subsection{Sufficient conditions for boundedness from below}

It is also useful to have simplified sufficient conditions which allow one to quickly determine if the potential is BFB for a given parameter set. Consider the top branch of \cref{necessaryandsufficient} --- i.e.\ $\Delta \geq 0$ and $-2 \leq \beta \leq 6$ and $\chi_1 \leq0$. One can show\footnote{Note that $\alpha = - \frac{1}{2} (\beta + 2)$ and $\gamma = - \frac{1}{2}(\beta + 2)$ are the directions along which $\Delta = 0$. In order to satisfy the $\chi_1 \leq 0$ condition, relevant for $\beta \leq 6$, we must have $\alpha+ \frac{1}{2} (\beta + 2) \geq 0$ and $\gamma + \frac{1}{2} (\beta + 2) \geq 0$. So long as $\beta \geq -2$, these conditions combined will always yield positive $\Delta$. Then \cref{BFBsufficientcondition1} is a sufficient condition that follows from the top branch of \cref{necessaryandsufficient}.} that a stronger condition (which will lead to a sufficient condition) is $\beta \leq 6$ \textit{and} $\alpha + \frac{1}{2} (\beta + 2) > 0$ and $\gamma + \frac{1}{2} (\beta + 2) > 0$. In terms of the $\lambda$'s, this translates to the sufficient condition:
\begin{equation}\label{BFBsufficientcondition1}
\begin{split}
    & 3 \sqrt{\lambda_1 \lambda_2} - (\lambda_3 + |\lambda_4| + |\lambda_5|) \geq 0 \,, \\
    & \text{and} \,\,\, \sqrt{\lambda_1 \lambda_2} + \lambda_3 - (|\lambda_4| + |\lambda_5| + 4 |\tilde{\lambda}_6|) > 0 \,, \\
    & \text{and} \,\,\, \sqrt{\lambda_1 \lambda_2} + \lambda_3 -( |\lambda_4| + |\lambda_5| + 4 |\tilde{\lambda}_7|) > 0 \,.
\end{split}
\end{equation}
Now consider the bottom branch of \cref{necessaryandsufficient} --- i.e.\ $\Delta \geq 0$ \textit{and} $\beta > 6$ \textit{and} $\chi_2 \leq0$. To arrive at an analytic sufficient condition, consider the stronger bound $\beta > 6$ \textit{and} $\alpha + 2 \sqrt{\beta -2} > 0$ \textit{and} $\gamma + 2 \sqrt{\beta -2} > 0$. In terms of the potential parameters, this condition reads:
\begin{equation}\label{BFBsufficientcondition2}
\begin{split}
    & \lambda_3 - (3 \sqrt{\lambda_1 \lambda_2} + |\lambda_4| + |\lambda_5|) \geq 0 \,, \\
    & \text{and} \,\,\, \sqrt{\lambda_3 - (\sqrt{\lambda_1 \lambda_2} + |\lambda_4| + |\lambda_5|) } - \frac{\sqrt{2}}{(\lambda_1 \lambda_2)^{1/4}} |\tilde{\lambda}_6| > 0 \,, \\
    & \text{and} \,\,\, \sqrt{\lambda_3 - (\sqrt{\lambda_1 \lambda_2} + |\lambda_4| + |\lambda_5|) } - \frac{\sqrt{2}}{(\lambda_1 \lambda_2)^{1/4}} |\tilde{\lambda}_7| > 0 \,.
\end{split}
\end{equation}
Eqs. (\ref{BFBsufficientcondition1}) and (\ref{BFBsufficientcondition2}) are the main result of this section.


\subsection{Numerical analysis}

\begin{figure}
    \centering
    \includegraphics[width=0.7\textwidth]{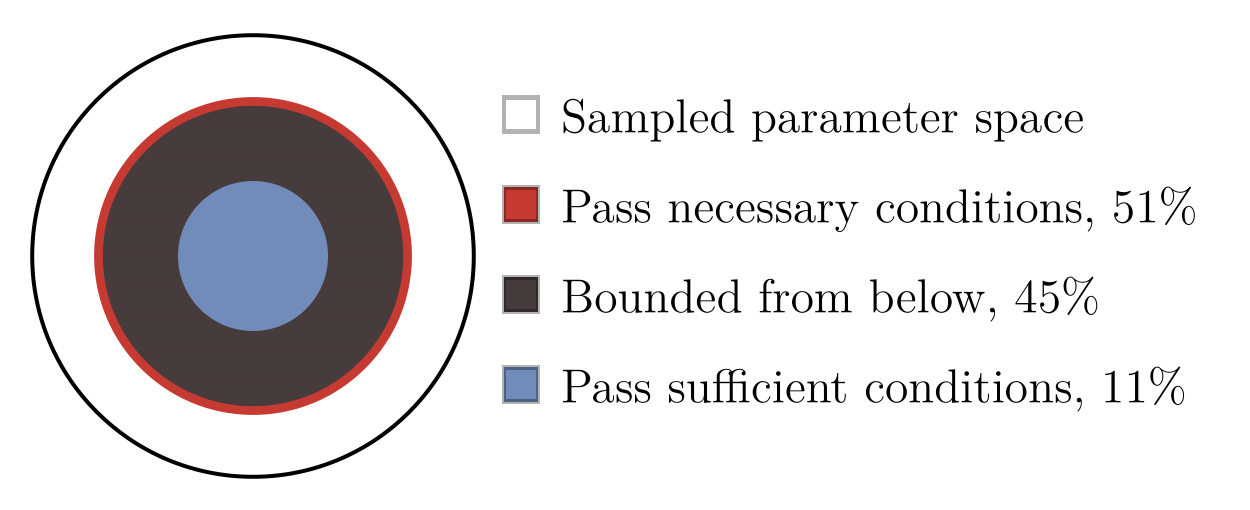}
    \caption{The white circle represents 10,000 randomly chosen points in the 7-dimensional parameter space of couplings $\{\lambda_1, ... \lambda_7\}$. We take as priors $\lambda_{1,2} \in [0,2\pi]$ and $|\lambda_{3,4,5,6,7}| \leq \frac{\pi}{2}$, with $\lambda_{5,6,7}$ allowed to be complex. The red circle encompasses the points which pass our analytic necessary conditions of Eqs. (\ref{eq:BFBnec2}), (\ref{eq:BFBnec3}), and (\ref{BFB_necessary}); the black circle contains the points which pass the necessary and sufficient BFB condition of \cref{necessaryandsufficient}; and the innermost blue circle contains the points which pass our sufficient condition of \cref{BFBsufficientcondition1}.}
    \label{fig:bfb_egg}
\end{figure}

To compare the performance of our analytic conditions with the numerical BFB condition, we perform a scan over 10,000 randomly chosen points in the 7-dimensional parameter space of $\{\lambda_1, ... \lambda_7\}$. We take as allowed ranges $\lambda_{1,2} \in [0,2\pi]$ and $|\lambda_{3,4,5,6,7}| \leq \frac{\pi}{2}$, with $\lambda_{5,6,7}$ complex, as this choice yields about half of the points BFB. \cref{fig:bfb_egg} shows the number of points which pass the numerical condition \cref{necessaryandsufficient} as well as the number which pass the combination of our necessary conditions Eqs.~(\ref{eq:BFBnec2}), (\ref{eq:BFBnec3}), and (\ref{BFB_necessary}) and the number which pass the combination of our sufficient conditions Eqs. (\ref{BFBsufficientcondition1}) and (\ref{BFBsufficientcondition2}). While we display in the figure the result of combining all necessary conditions derived in \cref{sec:BFB_necessary}, we note that \cref{eq:BFBnec3} provides the strongest necessary condition, with the combination of all conditions improving the results by a few percent.

We see that our necessary conditions are very effective at eliminating points which are not BFB, with only approximately 51\% of points passing this condition, as compared with the 45\% of points which actually satisfy BFB. Meanwhile, our analytic sufficient conditions guarantee approximately 11\% of points are BFB. Of these points, essentially all are obtained from \cref{BFBsufficientcondition1}, which was derived from the upper branch of \cref{necessaryandsufficient}. The second condition \cref{BFBsufficientcondition2}, derived from the lower branch, is too strong and admits almost no points, which also reflects the fact that most of the points sampled fall within the regime of the first branch.

\begin{figure}
    \centering
    \includegraphics[width=0.5\textwidth]{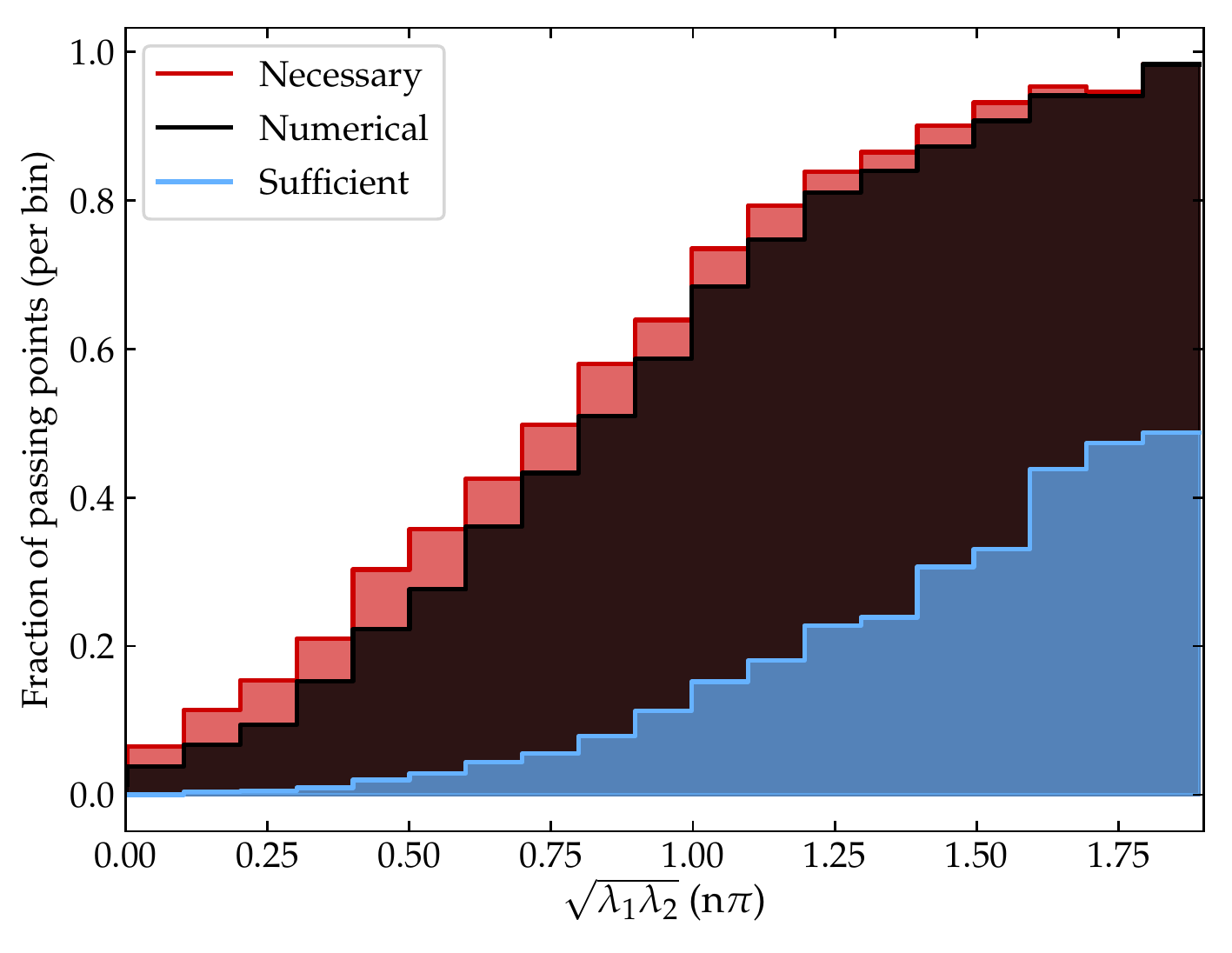}
    \caption{Histogram of $\sqrt{\lambda_1\lambda_2}$ displaying the fraction of tested points per bin which pass the necessary conditions Eqs.~(\ref{BFB_necessary}, \ref{eq:BFBnec2}, \ref{eq:BFBnec3}) (red), numerical test \cref{necessaryandsufficient} (black), and sufficient condition \cref{BFBsufficientcondition1} (blue). As in \cref{fig:bfb_egg}, we take as priors $\lambda_{1,2} \in [0,2\pi]$ and $|\lambda_{3,4,5,6,7}| \in [0, \frac{\pi}{2}]$, with $\lambda_{5,6,7}$ allowed to be complex.}
    \label{fig:bfb}
\end{figure}

An examination of the analytical expressions indicates that the quantity $\sqrt{\lambda_1\lambda_2}$ may play an important role in the determination of BFB. To examine whether the analytical form of our bounds indeed captures the primary underlying behavior of BFB with respect to the $\lambda_i$, we plot a histogram in $\sqrt{\lambda_1\lambda_2}$ of the fraction of tested points which pass the numerical, necessary, and sufficient bounds of Eqs.\ (\ref{necessaryandsufficient}),   (\ref{eq:BFBnec2}), (\ref{eq:BFBnec3}), (\ref{BFB_necessary}) and (\ref{BFBsufficientcondition1}), respectively. As in \cref{fig:bfb_egg}, we choose parameters in the range $\lambda_{1,2}\in[0,2\pi]$ and $|\lambda_{3,4,5,6,7}|\leq\pi/2$ with $\lambda_{5,6,7}$ allowed to be complex. The resulting figure is shown in \cref{fig:bfb}. As can be seen from the figure, more points pass the BFB condition for higher $\sqrt{\lambda_1\lambda_2}$, as indicated by the forms of the necessary and sufficient conditions. We find that both the necessary and sufficient bounds follow the same behavior as the exact numerical results, indicating that the analytic bounds do indeed capture the relevant behavior in $\sqrt{\lambda_1\lambda_2}$.

Finally, we note that within the existing literature, some simplified analytic BFB constraints for the most general 2HDM (i.e. involving $\lambda_{6,7} \neq 0$) do exist. For example, the authors of \ccite{Branco:2011iw} find as a necessary condition:
\begin{align}\label{BFB_literature}
     \frac{1}{2}(\lambda_1+\lambda_2) + \lambda_3 + \lambda_4 + \lambda_5 - 2 |\lambda_6 + \lambda_7| > 0 \,.
\end{align}
This expression, that agrees with  Eq.~(\ref{eq:BFBnec2}) in the appropriate limit, is derived by assuming that the Higgs doublets are aligned in field space, and is limited to the case that all $\lambda_i$ are taken to be real. Restricting ourselves to this regime, we find that the literature expression excludes approximately 17\% of points while ours excludes approximately 49\%, making our condition the stronger of the two by a large margin.

%% file: sec_stability.tex
We can also place constraints on the allowed 2HDM potential parameters by demanding the existence of a stable neutral vacuum. Strictly speaking, this not a necessary requirement: it is only necessary that the vacuum is meta-stable, with a lifetime longer than the age of the Universe. Here, we just derive the conditions for absolute stability, more precisely the absence of deeper minima at scales of the order of the TeV scale.

The discriminant $D$ introduced in \ccite{Barroso:2013awa,Ivanov:2015nea} offers a prescription for distinguishing the nature of a solution obtained by extremizing the potential. We summarize the method here, beginning by writing the potential as:
\begin{equation}\label{bilinearpotential}
	V = - M_\mu \mathsf{r}^\mu + \frac{1}{2} \Lambda_{\mu \nu} \mathsf{r}^\mu \mathsf{r}^\nu - \frac{1}{2} \zeta \mathsf{r}^\mu \mathsf{r}_\mu \,,
\end{equation}
where $M_\mu$ encodes the mass terms:
\begin{equation} \label{eq:Mmu}
	M_\mu = \begin{pmatrix} - \frac{1}{2} (m_{11}^2 + m_{22}^2) , & (m_{12}^2)^R, & - (m_{12}^2)^I, & - \frac{1}{2} (m_{11}^2 - m_{22}^2) \end{pmatrix} \,,
\end{equation}
$\mathsf{r}^\mu$ is a vector of field bilinears:
\begin{equation} \label{eq:rmu}
	\mathsf{r}^\mu = \begin{pmatrix} |\Phi_1|^2 + |\Phi_2|^2, & 2\text{Re}[\Phi_1^\dagger \Phi_2] , & 2\text{Im}[\Phi_1^\dagger \Phi_2] , & |\Phi_1|^2 - |\Phi_2|^2 \end{pmatrix} \,,
\end{equation}
and $\Lambda_{\mu \nu}$ encodes the quartic terms:
\begin{equation} \label{eq:Lambdamunu}
	\Lambda_{\mu \nu} = \frac{1}{2} \begin{pmatrix} \frac{1}{2}(\lambda_1 + \lambda_2) + \lambda_3 & \lambda_6^R + \lambda_7^R & - (\lambda_6^I + \lambda_7^I) & \frac{1}{2} (\lambda_1 - \lambda_2) \\ (\lambda_6^R + \lambda_7^R) & (\lambda_4 + \lambda_5^R) & - \lambda_5^I & \lambda_6^R - \lambda_7^R \\ -(\lambda_6^I + \lambda_7^I) & -\lambda_5^I & \lambda_4 - \lambda_5^R & -( \lambda_6^I - \lambda_7^I) \\ \frac{1}{2} (\lambda_1 - \lambda_2) & \lambda_6^R - \lambda_7^R & -(\lambda_6^I - \lambda_7^I )& \frac{1}{2}(\lambda_1 + \lambda_2) - \lambda_3 \end{pmatrix} \,.
\end{equation}
The last term in \cref{bilinearpotential} is a Lagrange multiplier we have introduced to enforce the condition $\mathsf{r}^\mu \mathsf{r}_\mu = 0$, which ensures we are in a charge-neutral minimum; we enforce this condition since charge-breaking and normal minima cannot coexist in the 2HDM (see \ccite{Barroso:2013awa,Ferreira:2004yd} for more details). In the above equations, indices are raised and lowered using a Minkowski metric. 

Provided the matrix $\Lambda_{\mu \nu}$, which contains the coefficients of the quartic terms in the potential, is positive definite, corresponding to a potential which is BFB, it can be brought into a diagonal form by an $SO(1,3)$ transformation:
\begin{equation}
	\Lambda_{\mu \nu}^{\text{diag}} = \begin{pmatrix} \Lambda_0 & 0 & 0 & 0 \\ 0 & - \Lambda_1 & 0 & 0 \\ 0 & 0 & - \Lambda_2 & 0 \\ 0 & 0 & 0 & - \Lambda_3 \end{pmatrix} \,,
\end{equation}
with $\Lambda_0$ the ``timelike'' eigenvalue and $\Lambda_i$ ``spacelike''. Let us define the ``signature matrix'' $S$ as $S \equiv \Lambda_{\mu \nu} - \zeta g_{\mu \nu}$. In diagonal form, it looks like:
\begin{equation}
	S =  \begin{pmatrix} \Lambda_0 - \zeta & 0 & 0 & 0 \\ 0 & \zeta - \Lambda_1 & 0 & 0 \\ 0 & 0 & \zeta - \Lambda_2 & 0 \\ 0 & 0 & 0 & \zeta - \Lambda_3 \end{pmatrix} \,.
\end{equation}
The discriminant is generically given by the determinant of the signature matrix:
\begin{equation}
	D = \det S \,.
\end{equation}
By using the diagonal form above, we can write this as:
\begin{equation}\label{diagonalD}
	D = (\Lambda_0 - \zeta)(\zeta - \Lambda_1)(\zeta - \Lambda_2)(\zeta - \Lambda_3) \,.
\end{equation}
We finally come to the vacuum stability condition. Suppose we have already verified that our potential is BFB and calculated the discriminant, time-like eigenvalue $\Lambda_0$, and Lagrange multiplier $\zeta$. 
\begin{equation}\label{globalmincondition}
    \text{We are in a global minimum if and only if}: \begin{cases} D>0 \,, \, \text{or} \\ D<0 \,\,\, \text{and} \,\,\, \zeta > \Lambda_0 \,. \end{cases}
\end{equation}

For our purposes, it is more useful to work with the ``Euclideanized" version of $\Lambda_{\mu \nu}$ obtained by lowering one of the indices with the Minkowski metric, $\Lambda_E \equiv \Lambda^\mu_\nu$. Explicitly:
\begin{equation}
	\Lambda_E = \frac{1}{2} \begin{pmatrix} \frac{1}{2}(\lambda_1 + \lambda_2) + \lambda_3 & \lambda_6^R + \lambda_7^R & - (\lambda_6^I + \lambda_7^I) & \frac{1}{2} (\lambda_1 - \lambda_2) \\ - (\lambda_6^R + \lambda_7^R) & - (\lambda_4 + \lambda_5^R) & \lambda_5^I & - (\lambda_6^R - \lambda_7^R) \\ \lambda_6^I + \lambda_7^I & \lambda_5^I & - (\lambda_4 - \lambda_5^R)& \lambda_6^I - \lambda_7^I \\ - \frac{1}{2} (\lambda_1 - \lambda_2) & -(\lambda_6^R - \lambda_7^R) & \lambda_6^I - \lambda_7^I & - \frac{1}{2}(\lambda_1 + \lambda_2) + \lambda_3 \end{pmatrix} \,.
\end{equation}
In terms of $\Lambda_E$, the discriminant is:
\begin{equation}
	D = - \det [\Lambda_E -\mathbb{1} \zeta ] \,.
\end{equation}

The other quantity necessary for formulating the discriminant is the Lagrange multiplier $\zeta$. This may be obtained by looking at any component of the minimization condition:
\begin{equation}
	\Lambda^\mu_\nu r^\nu - M^\mu = \zeta r^\mu \,.
\end{equation}
We parameterize the vacuum expectation values (vevs) of the doublets as:
\begin{equation}\label{generalvevs}
	\expval{\Phi_1} = \frac{1}{\sqrt{2}} \begin{pmatrix} 0 \\ v_1 \end{pmatrix} \,, \,\,\, \expval{\Phi_2} = \frac{1}{\sqrt{2}} \begin{pmatrix} 0 \\ v_2 e^{i \eta}\end{pmatrix} \,.
\end{equation}
Then the expectation value of field bilinears $r^\mu \equiv \expval{\mathsf{r}^\mu}$ is:
\begin{equation}
	r^\mu = \begin{pmatrix} \frac{1}{2} (v_1^2 + v_2^2), & v_1 v_2 \cos \eta , & v_1 v_2 \sin \eta , & \frac{1}{2} (v_1^2 - v_2^2) \end{pmatrix} \,.
\end{equation}
The expression for $\zeta$ is particularly simple if we choose the ``1'' component. In particular if we take $\eta = 0$, then:
\begin{equation}\label{zeta}
	\zeta = \frac{(m_{12}^2)^R}{v_1 v_2} - \frac{1}{2} \left( \frac{v_1}{v_2} \lambda_6^R + \frac{v_2}{v_1} \lambda_7^R + (\lambda_4 + \lambda_5^R) \right) \,.
\end{equation}
Note that this has the interpretation of the charged Higgs mass over the vev $v$ squared,
\begin{equation}
	\zeta = \frac{M_{H^\pm}^2}{v^2} \,,
\end{equation}
as first demonstrated in \ccite{Ivanov:2006yq,Maniatis:2006fs}. For a discussion of why this is generically true, see Appendix \ref{appA}.

If $D>0$, then the physical minimum is the global one, implying absolute stability. If, instead,  $D<0$, we need to compare the timelike eigenvalue $\Lambda_0$ with $\zeta$: we are in a global minimum if $\zeta > \Lambda_0$; otherwise, the minimum is metastable. Provided we have already verified that the potential is BFB, however, there is an even simpler way to assess the nature of the extremum. 

As an aside, working at the level of eigenvalues the two options of Eq. (\ref{globalmincondition}) for an extremum to be the global minimum can actually be collapsed into one. Recall that when the potential is BFB, $\Lambda^{\mu \nu}$ is positive definite and $\Lambda_0 > \Lambda_{1,2,3}$. Then from Eq. (\ref{diagonalD}), $D>0$ necessarily implies that we have the ordering $\Lambda_0 > \zeta > \Lambda_{1,2,3}$. Similarly $D<0$ and $\zeta > \Lambda_0$ necessarily implies the ordering $\zeta > \Lambda_0 > \Lambda_{1,2,3}$. So we see that the relative ordering of $\Lambda_0$ and $\zeta$ does not actually matter$-$all that matters for a potential which has been verified to be BFB is that $\zeta$ be larger than the spatial eigenvalues, $\zeta > \Lambda_{1,2,3}$.


\subsection{Sufficient conditions for stability}


\subsubsection{Gershgorin bounds}

As in \cref{sec:unitarity_sufficent}, we can bound the eigenvalues of $\Lambda_E$ using the Gershgorin disk theorem in order to derive a sufficient condition for a given vacuum solution to be stable. We first construct the intervals containing the eigenvalues of $\Lambda_E$ and define the endpoint of each interval as $\Gamma_i \equiv a_{ii} + R_i$, with $R_i = \sum_{j \neq i} |a_{ij}|$:
\begin{subequations}\label{gammas}
\begin{align}
	\Gamma_1 &= \frac{1}{4}(\lambda_1 + \lambda_2) + \frac{\lambda_3}{2} + \frac{1}{2} \left( |\lambda_6^R + \lambda_7^R | + | \lambda_6^I + \lambda_7^I | + \frac{1}{2} |\lambda_1 - \lambda_2| \right) \,, \\
	\Gamma_2 &=  -\frac{1}{4}(\lambda_1 + \lambda_2) + \frac{\lambda_3}{2} + \frac{1}{2} \left( |\lambda_6^R - \lambda_7^R | + | \lambda_6^I - \lambda_7^I | + \frac{1}{2} |\lambda_1 - \lambda_2| \right) \,, \\
	\Gamma_3 &= - \frac{1}{2} (\lambda_4 + \lambda_5^R) + \frac{1}{2} \bigg( |\lambda_5^I | + | \lambda_6^R + \lambda_7^R| + |\lambda_6^R - \lambda_7^R| \bigg) \,, \\
	\Gamma_4 &= - \frac{1}{2} (\lambda_4 - \lambda_5^R) + \frac{1}{2} \bigg( |\lambda_5^I | + | \lambda_6^I + \lambda_7^I | + |\lambda_6^I - \lambda_7^I | \bigg) \,.
\end{align}
\end{subequations}
We know that all eigenvalues must be less than the endpoint of the interval extending the furthest in the $+\hat{x}$ direction,
\begin{equation}
	 \text{max}[\Gamma_i] \geq \Lambda_{0,1,2,3} \,.
\end{equation}
Meanwhile, an extremum will be the global minimum if $\zeta> \Lambda_{1,2,3}$. Thus, it is sufficient to demand:
\begin{equation}\label{eq:vacuumGerschgorin}
	\zeta > \text{max}[\Gamma_i] \,.
\end{equation}


\subsubsection{Frobenius bounds}

One may also bound the eigenvalues using the Frobenius norm to obtain a single-equation condition. We require the maximum eigenvalue be less than $\zeta$, which gives the constraint:
\begin{align}\label{eq:vacuumFrobenius}
    \zeta > \frac{1}{2}\sqrt{\lambda_1^2 + \lambda_2^2 + 2(\lambda_3^2 + \lambda_4^2 + |\lambda_5|^2) + 4(|\lambda_6|^2 + |\lambda_7|^2)} 
\end{align}
Note that in this case, the Frobenius bound is insensitive to the signs of the $\lambda_i$, while the Gershgorin condition is sensitive to the signs of $\lambda_3$, $\lambda_4$, and $\lambda_5^R$. We thus expect the Gershgorin bound to be the stronger of the two.


\subsubsection{Principal minors}

In the case of a non-symmetric matrix such as $\Lambda_E$, Sylvester's criterion no longer holds and so cannot be applied in a straightforward manner. However, the following statement does hold: if the symmetric part of a matrix $M$ is positive-definite, then the real parts of the eigenvalues of $M$ are positive. This statement does not hold in the other direction, and therefore cannot be used to derive necessary conditions. However, we can apply Sylvester's criterion to the symmetric part of $\Lambda_E$ to obtain a sufficient condition.

The symmetric part of $\Lambda_E$ is given by
\begin{align}
    \Lambda_E^{S} &= \frac{1}{2}(\Lambda_E^{T} + \Lambda_E) \nonumber \\
    &= \frac{1}{2} \begin{pmatrix} 
        \frac{1}{2}(\lambda_1+\lambda_2) + \lambda_3 & 0 & 0 & 0 \\
        0 & -(\lambda_4 + \lambda_5^R) & \lambda_5^I & -(\lambda_6^R - \lambda_7^R) \\
        0 & \lambda_5^I & -(\lambda_4 - \lambda_5^R) & \lambda_6^I - \lambda_7^I \\
        0 & -(\lambda_6^R - \lambda_7^R) & \lambda_6^I - \lambda_7^I & -\frac{1}{2}(\lambda_1 + \lambda_2) + \lambda_3
        \end{pmatrix}
\end{align}
We require the matrix $\zeta \mathbb{1} - \Lambda_E^S$ to be positive-definite. Since the lower-right 3$\times$3 matrix decouples from the ``11'' element, we can analyze them separately when considering positive-definiteness. We require the ``11'' element to be positive, and apply Sylvester's criterion to the lower-right 3$\times$3 submatrix. This gives the following set of conditions:
\begin{subequations}\label{eq:vacuumSylvester}
\begin{equation}
    \zeta - \frac{1}{4}(\lambda_1 + \lambda_2) - \frac{1}{2}\lambda_3 > 0 \,,
\end{equation}
\begin{equation}
    \zeta + \frac{1}{2}\lambda_4 + \frac{1}{2}\lambda_5^R > 0 \,,
\end{equation}
\begin{equation}
    \big(\zeta + \frac{1}{2}\lambda_4\big)^2 - \frac{1}{4}|\lambda_5|^2 > 0 \,,
\end{equation}
\begin{equation}
\begin{split}
    (4\zeta + \lambda_1 + \lambda_2 & -2 \lambda_3)((2 \zeta + \lambda_4)^2 - |\lambda_5|^2)\\ & + \lambda_5 (\lambda_6^* - \lambda_7^*)^2 + \lambda_5^* (\lambda_6 - \lambda_7)^2 - 2 (2 \zeta + \lambda_4) |\lambda_6 - \lambda_7|^2 >0 \,.
\end{split}
\end{equation}
\end{subequations}
Taken together, the Eqs.~(\ref{eq:vacuumSylvester}) provide a sufficient condition for vacuum stability.


\subsection{Numerical comparison}

\begin{figure}
    \centering
    \includegraphics[width=0.6\textwidth]{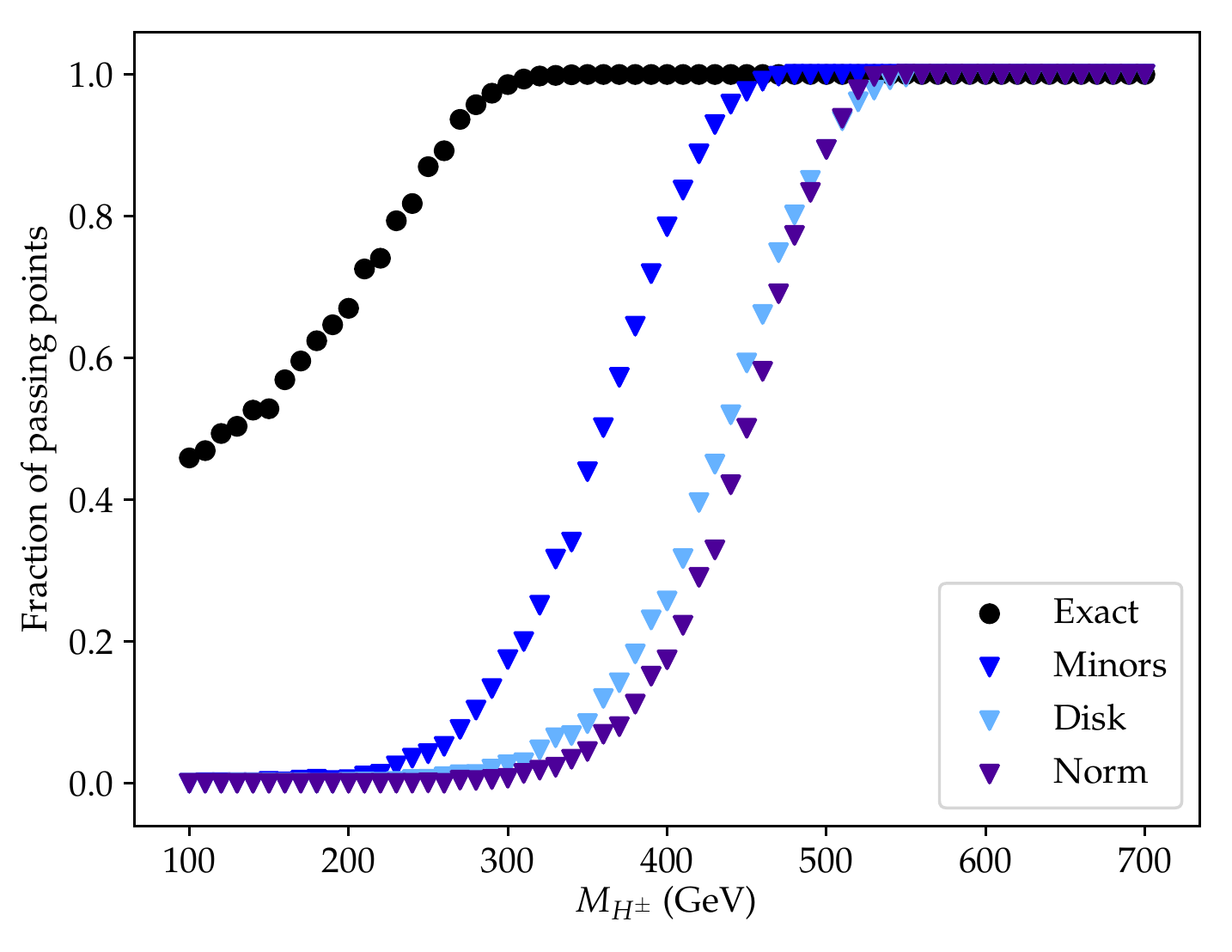}
    \caption{Comparison of the fraction of points that pass the exact stability conditions, Eq.~(\ref{globalmincondition}) (black dots), with respect to the fraction passing the three sufficient conditions for vacuum stability: principal minors \cref{eq:vacuumSylvester} (blue), Gershgorin disk theorem \cref{eq:vacuumGerschgorin} (light blue), and Frobenius norm \cref{eq:vacuumFrobenius} (purple). We plot the fraction of points that pass each condition as a function of $M_{H^{\pm}}$. We take $\lambda_{1,2}\in[0,2\pi]$ and $|\lambda_{3,4,5,6,7}|\leq\frac{\pi}{2}$, with $\lambda_{5,6,7}$ allowed to be complex, and restrict to examining points which are BFB. }
    \label{fig:vacuum_stab}
\end{figure}

In \cref{fig:vacuum_stab} we plot the performance of the three sufficient conditions for vacuum stability, Eqs.  (\ref{eq:vacuumGerschgorin}), (\ref{eq:vacuumFrobenius}), (\ref{eq:vacuumSylvester}), as a function of the charged Higgs mass $M_{H^{\pm}}$. We compare these results with the fraction that pass the exact stability condition, \cref{globalmincondition}. As in previous sections, we choose the $\lambda_i$ randomly with $\lambda_{1,2}\in[0,2\pi]$ and $|\lambda_{3,4,5,6,7}|\leq\frac{\pi}{2}$, with $\lambda_{5,6,7}$ allowed to be complex. The y-axis shows the fraction of tested points which pass the respective stability condition; we restrict to testing points which are BFB, to ensure the validity of the stability conditions implemented in ~\cref{fig:vacuum_stab}. We find that the set of conditions arising from the application of Sylvester's criterion capture the most stable points, while all three bounds capture more stable points when the $\lambda_{i}$ are small compared to the ratio $M_{H^\pm}^2/v^2$.


\subsection{Vacuum stability in the Higgs basis}

It is particularly interesting to study vacuum stability in the Higgs basis, in which only one of the doublets possesses a vev (see \cref{Higgsbasisconversion} for a review of the conversion to the Higgs basis as well as our conventions). One advantage of this basis is that the potential parameters are closely related to physical observables: for example, $Z_1$ controls the trilinear coupling of three SM-like Higgs bosons $hhh$, $Z_6$ controls the trilinear coupling of two SM-like and one non-SM-like \cp-even Higgs bosons $hhH$, etc. (see e.g.~\ccite{Carena:2015moc} for an exhaustive list of couplings). 
Since none of the bounds obtained in this article have relied on the choice of basis, they can equally well be applied to Higgs basis parameters. Using the close relationship between the Higgs basis parameters and physical quantities, we here aim at obtaining approximate bounds on the physical observables of the model.

In our notation, given in \cref{Higgsparam} of Appendix A (choosing $\eta = 0$), the scalar which obtains a vev is denoted by $\phi_1^0$. The mass matrix for the neutral scalars $\vec{\phi} = (\phi_1^0, \phi_2^0, a_0)^{\text{T}}$ reads:
\begin{equation}\label{neutralmass}
    \mathcal{M}^2 = v^2 \begin{pmatrix} Z_1 & Z_6^R & - Z_6^I \\ Z_6^R & \frac{M_{H^\pm}^2}{v^2}+ \frac{1}{2} (Z_4 + Z_5^R) & - \frac{1}{2} Z_5^I \\ - Z_6^I & - \frac{1}{2} Z_5^I & \frac{M_{H^\pm}^2}{v^2} + \frac{1}{2} (Z_4 - Z_5^R) \end{pmatrix} \,,
\end{equation}
where $M_{H^\pm}^2$ is the charged Higgs mass:
\begin{equation}\label{chargedH}
    M_{H^\pm}^2 = M_{22}^2 + \frac{1}{2} Z_3 v^2 \,.
\end{equation} 

We will restrict ourselves to the alignment limit, which is the limit in which $\phi_1^0$ is aligned with the 125 GeV mass eigenstate. In this case, the 125 GeV Higgs couples to the electroweak gauge bosons and all fermions with SM strength, and the alignment limit is therefore phenomenologically well-motivated by precision Higgs results from the LHC~\cite{CMS:2018uag,ATLAS:2019nkf}. 

Examining the above matrix, it appears that there are two ways in which one may obtain alignment. The first option, known as the decoupling limit, corresponds to taking $M_{H^\pm}^2 + \frac{1}{2} (Z_4 \pm Z_5^R)v^2 \gg Z_1 v^2$. Under this limit, the heavy mass eigenstates $h_2$ and $h_3$ and the heavy charged Higgs $H^\pm$ decouple from the light mass eigenstate, leaving $h_1$ aligned with $\phi_1^0$. More interesting from a phenomenological standpoint is the approximate \textit{alignment without decoupling} limit, as it leaves the non-standard Higgs states potentially within collider reach. This corresponds to taking $|Z_6| \ll 1$, for which the mixing between $\phi_1^0$ and the other neutral scalars vanishes, leading  to the identification of $\phi_1^0$  with the mass eigenstate $h_1$. For the following discussion we will take $|Z_6| \ll 1$ and work in the alignment without decoupling limit.

We define $h_1 \equiv h$ to be the SM-like Higgs boson, which has a mass given by
\begin{equation}\label{approxSMhiggs}
    M_{h}^2 = Z_1 v^2 \,.
\end{equation}
To obtain a physical Higgs mass close to the experimental value of 125 GeV, it is required that we fix $Z_1 \approx 0.25$. The remaining 2$\times$2 mass matrix can be diagonalized to obtain the masses of the remaining scalars $h_2$ and $h_3$:
\begin{equation}
    M_{h_3,h_2}^2 = M_{H^\pm}^2 + \frac{1}{2} (Z_4 \pm |Z_5|) v^2 \,.
    \label{eq:Mh2h3}
\end{equation}
There are two possibilities for the \cp properties of these states. So long as $Z_5^I \neq 0$, $h_2$ and $h_3$ have mixed \cp properties. In the limit of $Z_5^I = 0$, meanwhile, the non-standard Higgs mass matrix becomes diagonal, and we obtain mass eigenstates $H$ and $A$ with definite \cp character,
\begin{equation}
    M_{H,A}^2 = M_{H^\pm}^2 + \frac{1}{2} (Z_4 \pm Z_5^R) v^2 \,.
    \label{eq:MHA}
\end{equation}  
The masses of the general mass eigenstates and the states of definite \cp character can be related by 
\begin{eqnarray}
M_{H}^2 & = & M_{h_3,h_2}^2 + \frac{1}{2} ( Z_5^R \mp |Z_5| )v^2 
\nonumber\\
M_A^2  & = & M_{h_3,h_2}^2 - \frac{1}{2} ( Z_5^R \pm |Z_5| )v^2 .
\end{eqnarray}
In the following analysis we will make no assumptions about the \cp character of the mass eigenstates, and will work with the generic physical masses $M_{h_3,h_2}$.

With the above definitions, we can rephrase our sufficient vacuum stability conditions into constraints on physical quantities. We start with the Gershgorin condition of \cref{eq:vacuumGerschgorin}. Expressing the $\Gamma$'s in terms of physical masses, a sufficient condition for vacuum stability becomes:
\begin{equation}
\begin{split}
    & M_{h_2}^2 > \frac{1}{2} (|Z_5^I| - Z_5^R - |Z_5|) v^2 + \frac{1}{2} |Z_7^R| v^2 \,\,\, \text{and}\\
    & M_{h_3}^2  > \frac{1}{2} (|Z_5^I| + Z_5^R + |Z_5|) v^2 + \frac{1}{2} |Z_7^I| v^2 \,\,\,\,\, \text{and}\\
    & M_{H^\pm}^2 > \frac{1}{2}\, \text{max}[M_h^2, \, Z_2 v^2] + \frac{1}{2} (Z_3 + |Z_7^R| + |Z_7^I|)v^2 \,.
\end{split}
\end{equation}

Next, we can recast the Frobenius sufficient condition, \cref{eq:vacuumFrobenius}, in terms of the physical masses; doing so results in the following condition:
\begin{equation}
    2M_{H^\pm}^2 \big(M_{h_2}^2 + M_{h_3}^2 \big) - \big(M_{H^\pm}^4 + M_{h_2}^4 + M_{h_3}^4 + \frac{1}{4} M_h^4 \big) > \frac{1}{4} (Z_2^2 + 2 Z_3^2 + 4 |Z_7|^2) v^4 \,.
\end{equation}
Finally, Sylvester's criterion provides an additional set of sufficient conditions. A sample set of sufficient conditions for vacuum stability in the alignment limit based on \cref{eq:vacuumSylvester} is:
\begin{equation}
\begin{split}
    & 4 M_{H^\pm}^2 - M_h^2 > (Z_2 + 2 Z_3)v^2, \hspace{0.75cm} \\
    & M_{h_3}^2 > \frac{1}{2}(|Z_5|- Z_5^R), \hspace{4.cm} \\
    &M_{h_2}^2 M_{h_3}^2 > 0, \hspace{3.6cm} \\
    4 M_{h_2}^2 M_{h_3}^2 ( 4 M_{H^\pm}^2 + M_h^2 + Z_2 v^2 &-2 Z_3 v^2) - 2 (M_{h_2}^2 + M_{h_3}^2) |Z_7|^2 + Z_5 {Z_7^*}^2 + Z_5^* Z_7^2 > 0 \,.
\end{split}
\end{equation}

%% file: sec_CP.tex
The bounds we have derived in this work have implications for the allowed values of physical parameters in a given 2HDM. This can be seen in a straightforward way in the previous section, where the conditions for vacuum stability were recast into expressions that restrict the physical masses of the bosonic sector. One particularly interesting question to which our bounds can be applied is that of the amount of \cp violation permitted in the alignment limit. This possibility has largely been neglected in the many previous studies which restrict themselves to the $\mathbb{Z}_2$-symmetric 2HDM. This is understandable since exact alignment implies \cp conservation in the $\mathbb{Z}_2$-symmetric case. When working in the fully general 2HDM, however, it is possible to have \cp violation whilst still keeping the SM-like Higgs boson fully aligned.

To justify this claim, recall that there are four complex parameters in the 2HDM: $\{ M_{12}^2, Z_5, Z_6, Z_7 \}$. One of these is fixed by the minimization condition $M_{12}^2 = - \frac{1}{2} Z_6 v^2$, leaving just three independent parameters, which we take to be the couplings $\{ Z_5, Z_6, Z_7 \}$. These complex parameters enter into the three basic \cp violating invariants of the 2HDM scalar sector $J_1$, $J_2$, and $J_3$, which can be thought of as analogous to the Jarlskog invariant $J$ of the SM quark sector. They are worked out explicitly in ~\cite{Lavoura:1994fv,Davidson:2005cw}; the important fact is that they scale as $J_1 \sim \text{Im}[Z_5^* Z_6^2]$, $J_2 \sim \text{Im}[Z_5^* Z_7^2]$, $J_3 \sim \text{Im}[Z_6^* Z_7]$. It is then clear that the condition for the Higgs sector to be \cp invariant is:
\begin{equation}\label{CPCcondition}
    \text{Im}[Z_5^* Z_6^2] = \text{Im}[Z_5^* Z_7^2] = \text{Im}[Z_6^* Z_7] = 0 \,.
\end{equation}
There are two ways in which this can be satisfied~\cite{Low:2020iua}: either $Z_5^I = Z_6^R = Z_7^R = 0$ or $Z_5^I = Z_6^I = Z_7^I = 0$. Note that in the limit of exact alignment we have $Z_6 = 0$, so this reduces to demanding either $Z_5^I = Z_7^R = 0$ or $Z_5^I = Z_7^I = 0$.


Meanwhile in the 2HDM with a (softly broken) $\mathbb{Z}_2$ symmetry, the fact that $\lambda_6 = \lambda_7 = 0$ implies the following two relations between the parameters in the Higgs basis (see Eq.~(\ref{quarticconversions}) in Appendix A)~\cite{Haber:2015pua,Boto:2020wyf}
\begin{align}
    Z_6 + Z_7 &= \frac{1}{2}\tan 2\beta (Z_2 - Z_1), \\
    Z_6 - Z_7^* &= \frac{1}{\tan 2\beta}( Z_1 + 2 Z_6 \cot 2\beta - Z_3 - Z_4 - Z_5).
\end{align}
It immediately follows that:
\begin{align}
    Z_7^I &= - Z_6^I, \\
    Z_5^I &= 2 \; \frac{1 - \tan^2 2\beta}{\tan 2\beta}Z_6^I.\label{Z5i_Z6i_relation}
\end{align}
These conditions imply that in the exact alignment limit (i.e. $Z_6 = 0$), it will necessarily be the case that $Z_5^I = Z_7^I = 0$. Thus, exact alignment directly leads to \cp conservation in the $\mathbb{Z}_2$-symmetric or softly broken $\mathbb{Z}_2$-symmetric 2HDM. This need not be the case in the fully general 2HDM, where the above relations no longer hold. If we allow for a small misalignment (i.e. $|Z_6| \gtrsim 0$), |$Z_5^I$|, which controls the mixing between the $H$ and $A$ bosons (see \cref{neutralmass}), can still be large for large $\tan\beta$.

\begin{figure}
    \centering
    \includegraphics[width=0.49\textwidth]{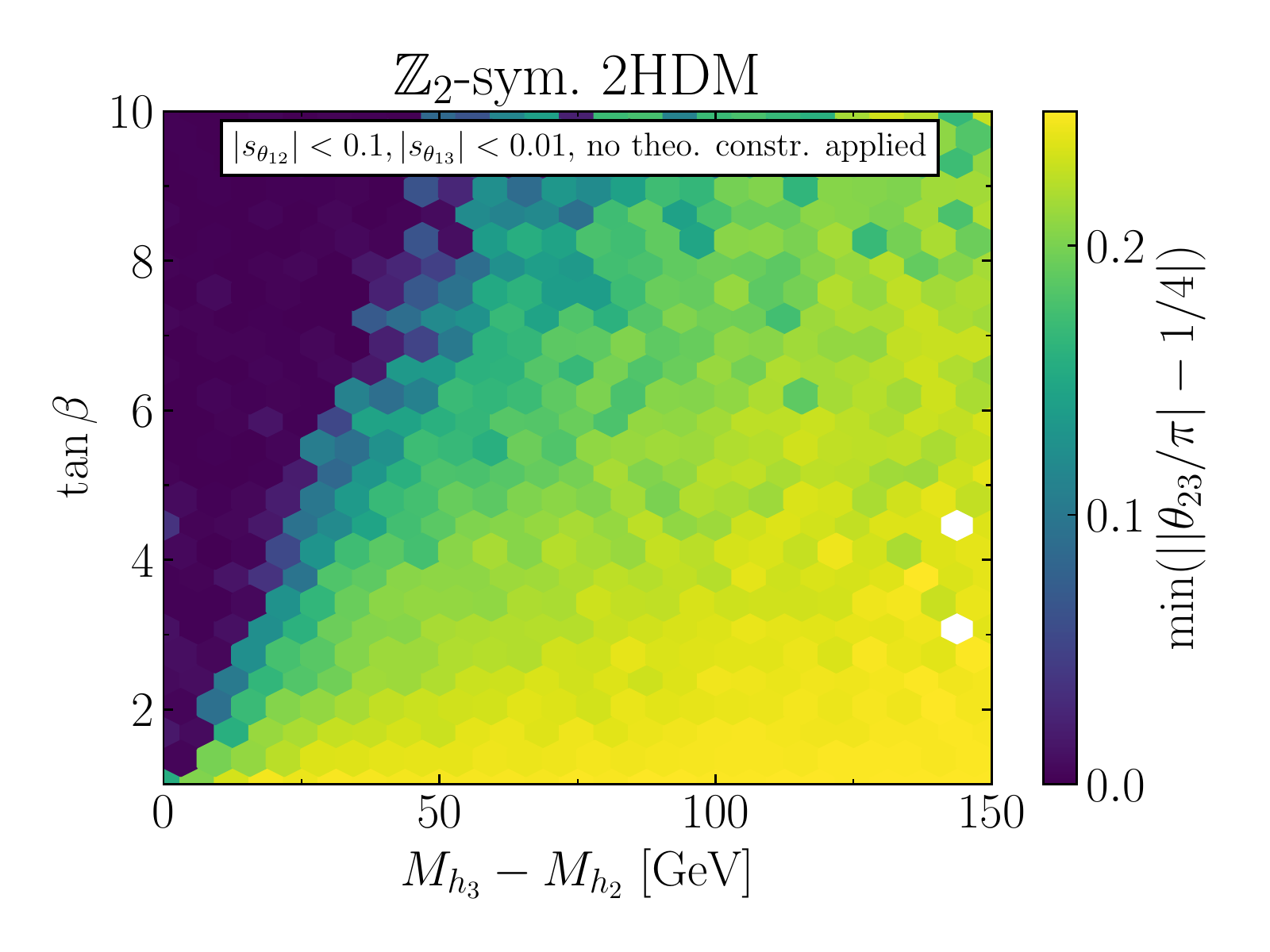}
    \includegraphics[width=0.49\textwidth]{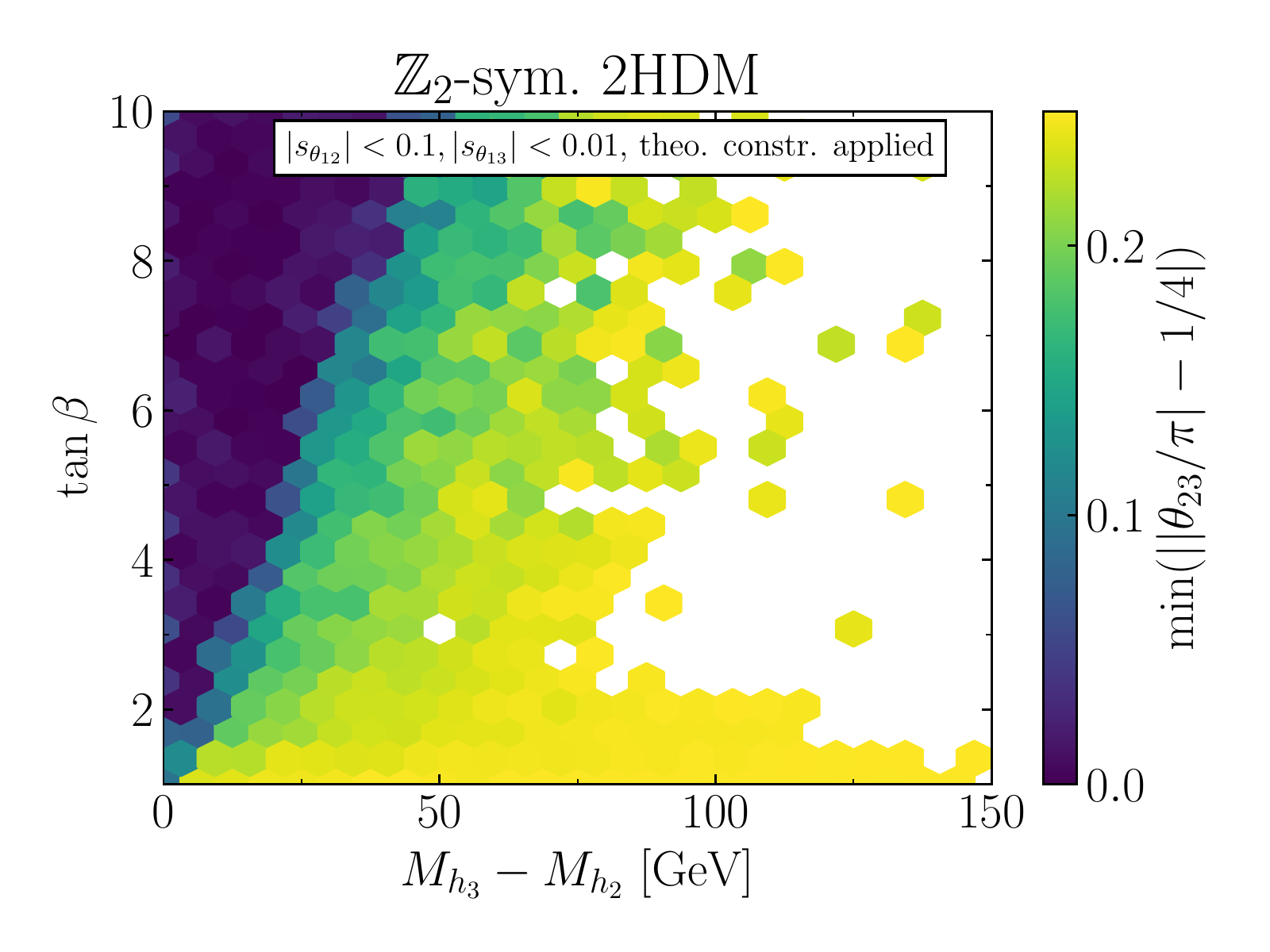}
    \includegraphics[width=0.49\textwidth]{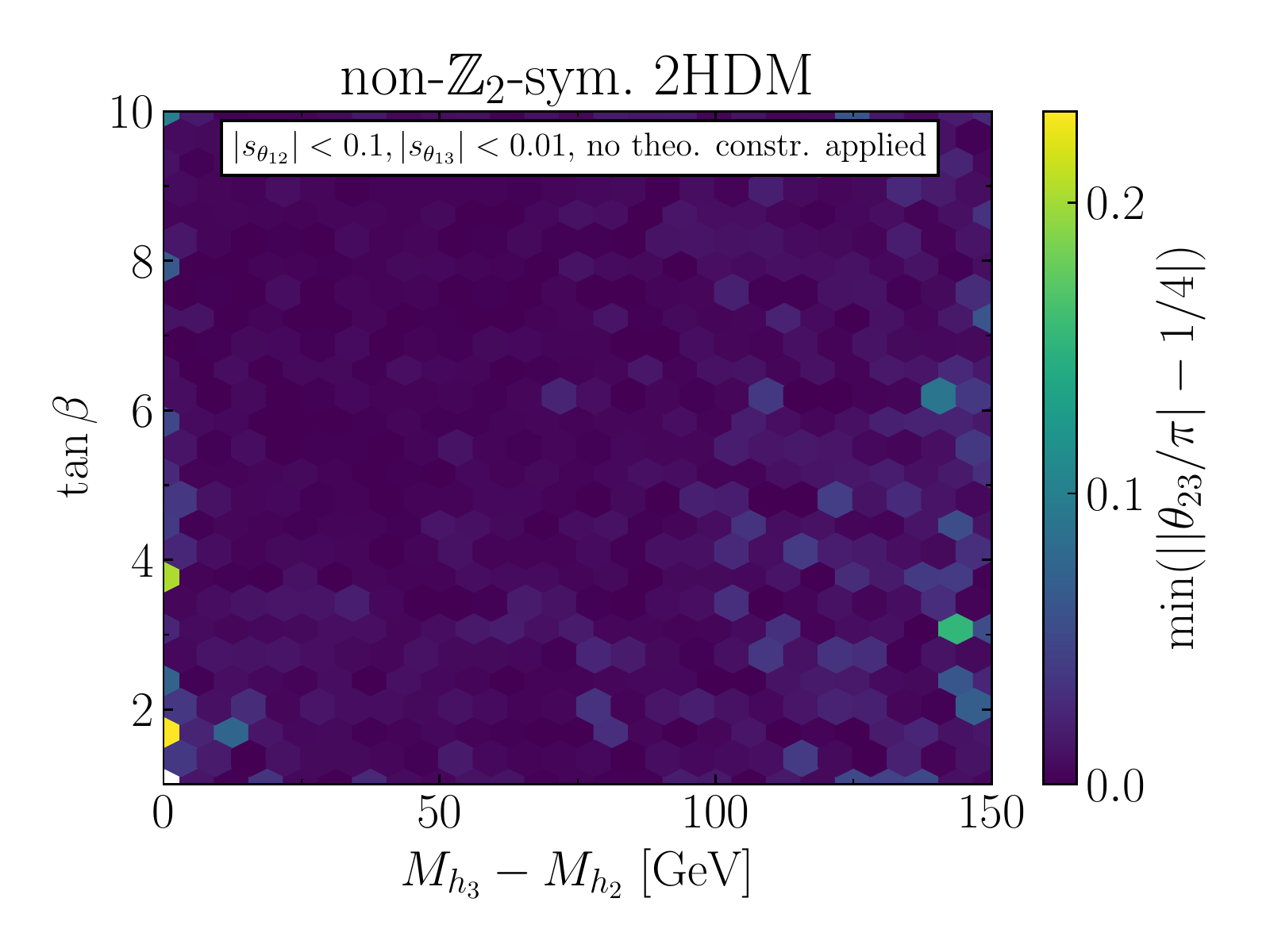}
    \includegraphics[width=0.49\textwidth]{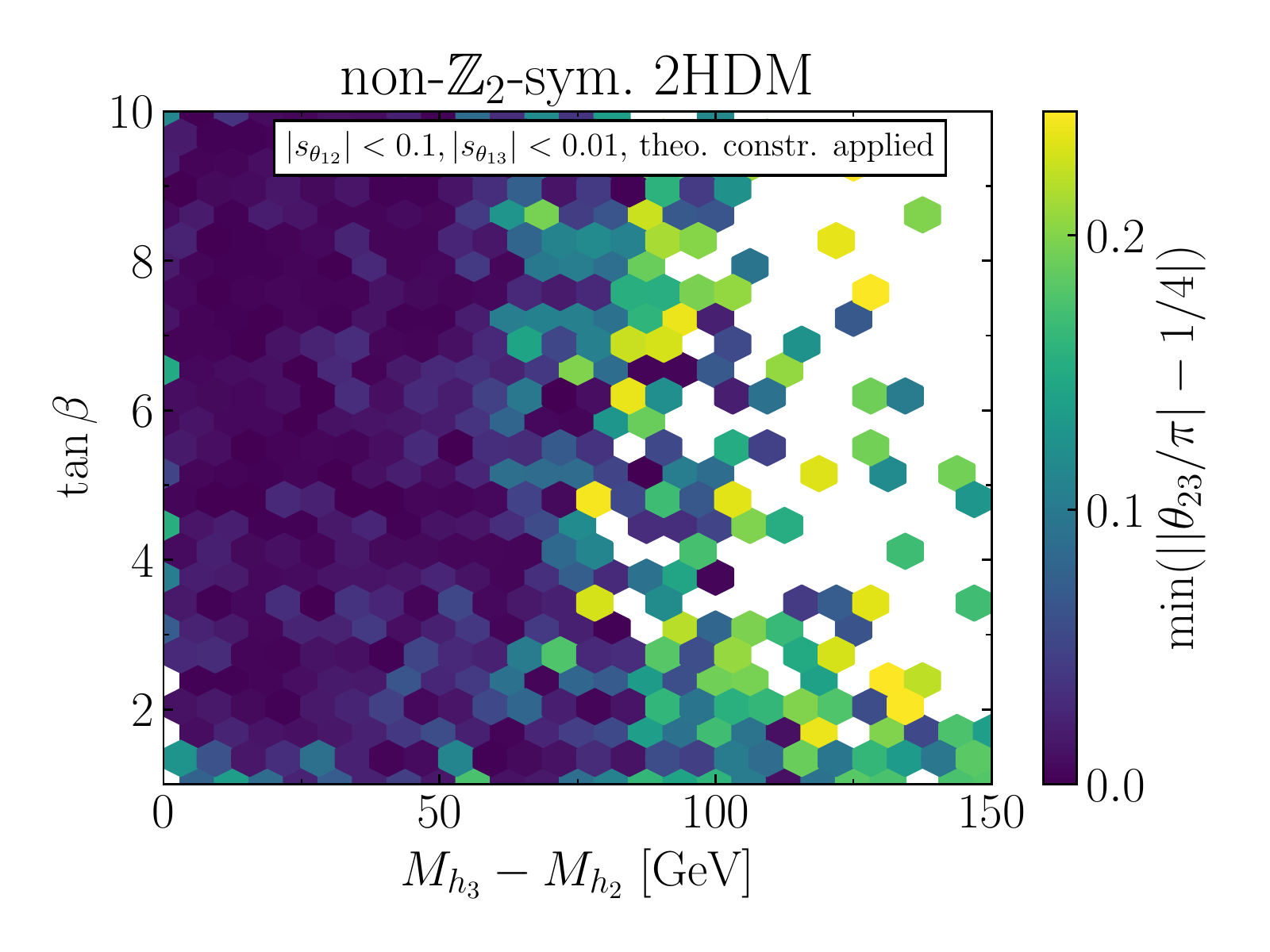}
    \caption{\textit{Upper left:} Parameter scan of the 2HDM with a softly broken $\mathbb{Z}_2$ symmetry in the $(M_{h_3} - M_{h_2},\tan\beta)$ parameter plane with parameter range $\lambda_{1,2}\in [0,2\pi]$ and $|\lambda_{3,4,5}|\in[0,\pi]$. The conditions $|\sin\theta_{12}| < 0.1$ and $|\sin\theta_{13}| < 0.01$ are imposed. The colour indicates the minimal value of $|\theta_{23}/\pi - 1/4|$ in each hexagonal patch. \textit{Upper right:} Same as upper left, but constraints from perturbative unitary, BFB, and vacuum stability are applied in addition. \textit{Lower left:} Same as upper left, but the scan is performed in the general 2HDM without a (softly broken) $\mathbb{Z}_2$ symmetry ($|\lambda_{6,7}|\in[0,\pi]$). \textit{Lower right:} Same as lower left, but constraints from perturbative unitary, BFB, and vacuum stability are applied in addition.}
    \label{fig:CP_mixing}
\end{figure}

The physical consequences of the difference in the \cp properties in the alignment limit between the (softly broken) $\mathbb{Z}_2$-symmetric and the general 2HDMs are illustrated in \cref{fig:CP_mixing}, in which we show the results of several parameter scans. Motivated by Higgs precision~\cite{CMS:2018uag,ATLAS:2019nkf} and electric dipole moment bounds~(see e.g.\ \ccite{ACME:2018yjb,Altmannshofer:2020shb}), we demand a small mixing of the $h$ and the $H$ states and an even smaller mixing of the $h$ and the $A$ states; we impose these constraints on the mixing in an approximate manner by demanding $|\sin\theta_{12}| < 0.1$ and $|\sin\theta_{13}| < 0.01$, where $\theta_{12}$ and $\theta_{13}$ are the respective mixing angles.\footnote{The stated bounds on $\theta_{12}$ and $\theta_{13}$ are just illustrative. We checked that our conclusions do not depend strongly on the values inserted.} The colour code in the figure indicates the minimum value of $|\theta_{23}/\pi - 1/4|$ in each hexagonal patch, where $\theta_{23}$ is the mixing angle between the $H$ and $A$ states. This variable is chosen such that the maximal mixing case corresponds to a value of zero. Correspondingly, a dark blue color indicates that a large \cp-violating mixing between the $H$ and $A$ states can be realized; a bright yellow color instead signals that no large mixing can be realized.

In the upper left panel of \cref{fig:CP_mixing} we present a parameter scan of the 2HDM with a softly broken $\mathbb{Z}_2$ symmetry in the $(M_{h_3} - M_{h_2},\tan\beta)$ parameter plane. We observe that for large $\tan\beta$, a large mixing between the neutral BSM Higgs bosons can be realized if their mass difference is below $\sim 70\gev$. Larger mass differences originate from differences between the diagonal terms of the Higgs mass matrix (see \cref{neutralmass}), suppressing possible mixing effects induced by the off-diagonal $\frac{1}{2} Z_5^I v^2$ term. For lower $\tan\beta$, the condition $|\sin\theta_{13}|<0.01$ directly implies that $Z_5^I$ is small (see \cref{Z5i_Z6i_relation}), resulting in substantial mixing only when the mass difference is close to zero.

In the upper right panel of \cref{fig:CP_mixing}, we again consider a softy broken $\mathbb{Z}_2$ symmetry but additionally impose perturbative unitarity, boundedness-from-below, and vacuum stability constraints following the discussions in the previous Sections. Aside from lowering the maximal possible mass difference between $h_3$ and $h_2$, the region in which large mixing between the BSM Higgs bosons can be realized is also reduced.

If we instead investigate the general 2HDM without a (softly broken) $\mathbb{Z}_2$ symmetry (see lower panels of \cref{fig:CP_mixing}), large mixing between the $H$ and $A$ states can be realized throughout the shown parameter plane. Applying the bounds derived in the previous sections only excludes the region with $M_{h_3} - M_{h_2} \gtrsim 100\gev$.

Finally, we want to remark that \cp violation can become manifest not only in the neutral mass matrix but also in the bosonic couplings. This occurs if either $\tilde{Z}_7^R \neq 0$ or $\tilde{Z}_7^I \neq 0$, since these couplings enter into couplings like $g_{h_1 h_2 h_3}$~\cite{Low:2020iua}. Exotic decays like $h_3 \rightarrow h_1 h_2 \rightarrow 3 h_1$ would then be indicative of \cp violation in the bosonic sector (see e.g.~\cite{Haber:2022gsn}).

%% file: sec_conclusions.tex
Two Higgs doublet models (2HDMs) present a natural extension of the Standard Model description. In spite of the simplicity of this SM extension, many new parameters appear in this theory, and it is very important to understand the constraints on these parameters which will impact in a relevant way the 2HDM phenomenology. Most existing studies concentrate on the case in which a $\mathbb{Z}_2$ symmetry is imposed on the 2HDM potential and Yukawa sector. While this symmetry is an easy way to avoid flavor-changing neutral currents, it also forbids certain terms in the Higgs potential which do not induce flavor-changing neutral currents at tree level. In fact, in many scenarios in which the 2HDM is the low-energy effective field theory of a more complete high-scale model, these couplings are predicted to be non-zero.

\begin{table}
\begin{tabular}{|c||c|c|c|}
 \hline 
 Conditions & Perturbative unitarity & Bounded from below & Vacuum stability \\ 
 \hline\hline
 Exact & \cref{eq:unitarity_exact} & \cref{necessaryandsufficient} & \cref{globalmincondition} \\ 
 \hline
 Necessary & \cref{eq:unitarity_D2,eq:unitarity_D3} & \cref{eq:BFBnec3} & --- \\
 \hline
 Sufficient & \cref{eq:unitarityFrobenius} & \cref{BFBsufficientcondition1} &  \cref{eq:vacuumSylvester} \\
 \hline 
\end{tabular}
\caption{Overview of the primary results of this paper; further constraints and their analysis may be found in the main text.}
\label{tab:overview}
\end{table}

Based on this motivation, in this work we present a step towards a systematic exploration of the non-$\mathbb{Z}_2$-symmetric 2HDM. We studied three of the most important theoretical constraints on the scalar potential parameters: perturbative unitarity, boundedness from below, and vacuum stability. In all three cases, we concentrated on the most general renormalizable potential (not restricted by any discrete symmetry) extending previous works by deriving analytic necessary and sufficient conditions for these constraints. For convenience, our main results (i.e., those conditions which approximate the exact conditions the best) are summarized in \cref{tab:overview}.

The derivation of our constraints makes use of several relevant mathematical properties, of which many have not been exploited in the literature before. These properties are not only applicable to the 2HDM but are also useful for the exploration of other models with extended Higgs sectors.

As a first phenomenological application of our bounds, we studied how much \cp-violating mixing between the BSM Higgs bosons can be realized in the general 2HDM in comparison to a 2HDM with a (softly broken) $\mathbb{Z}_2$ symmetry. While we found that large \cp-violating mixing can only be realized for large $\tan\beta$ in the 2HDM with a softly broken $\mathbb{Z}_2$ symmetry, no such theoretical constraints exist for the general 2HDM.

We leave a comprehensive study of the phenomenological consequences of not imposing a $\mathbb{Z}_2$ symmetry for future work.

%% file: appendix.tex
\section{Higgs basis conversion}\label{Higgsbasisconversion}

The phenomenological properties of the Higgs sector are more easily analyzed in the Higgs basis, in which only one of the doublets possesses a vev\footnote{This is technically not enough to uniquely define the Higgs basis. The $U(1)$ diagonal subgroup of the $SU(2)$ symmetry in Higgs flavor space remains intact following SSB. This corresponds to transformations $\Phi_1 \rightarrow e^{i\chi} \Phi_1$, $\Phi_2 \rightarrow e^{-i \chi} \Phi_2$. As a result, we have a one-dimensional family of Higgs bases parameterized by $\chi$: $\{e^{-i\chi} H_1, e^{i\chi} H_2\}$.}. We parameterize the doublets as:
\begin{equation}\label{Higgsparam}
	H_1 = \begin{pmatrix} G^+ \\ \frac{1}{\sqrt{2}}(v+ \phi_1^0 + i G^0) \end{pmatrix} \,, \,\,\, H_2 = \begin{pmatrix} H^+ \\ \frac{1}{\sqrt{2}}(\phi_2^0 + i a^0) \end{pmatrix} \,,
\end{equation}
where $G^\pm$ and $G^0$ are the Goldstones that become the longitudinal components of $W^\pm$ and $Z$, $H^\pm$ is the physical singly charged scalar state, and $(\phi_1^0, \phi_2^0, a^0)$ are the neutral scalars. The potential in the Higgs basis reads:
\begin{equation}\label{Higgsbasispotential}
\begin{split}
	V & = M_{11}^2 H_1^\dagger H_1 + M_{22}^2 H_2^\dagger H_2 - (M_{12}^2 H_1^\dagger H_2 + h.c.)\\
	& + \frac{1}{2} Z_1 (H_1^\dagger H_1)^2 + \frac{1}{2} Z_2 (H_2^\dagger H_2)^2 + Z_3 (H_1^\dagger H_1)(H_2^\dagger H_2) + Z_4 (H_1^\dagger H_2) (H_2^\dagger H_1)\\
	& + \left[ \frac{1}{2} Z_5 (H_1^\dagger H_2)^2 + Z_6 (H_1^\dagger H_1)(H_1^\dagger H_2) + Z_7 (H_2^\dagger H_2) (H_1^\dagger H_2) + h.c. \right] \,.
\end{split}
\end{equation}
The conversion between the potential parameters in the weak eigenstate basis and those in the Higgs basis have been worked out in \ccite{Davidson:2005cw};  to be self-contained, we reproduce them here. They are obtained by a rotation by an angle $\beta$ in field space of the original two Higgs doublets. The mass terms in the two bases are related as:
\begin{subequations}
\begin{align}
    m_{11}^2 &= M_{11}^2 c_\beta^2 + M_{22}^2 s_\beta^2 + \text{Re}[M_{12}^2 e^{i \eta}] s_{2\beta} \,, \\
    m_{22}^2 &= M_{11}^2 s_\beta^2 + M_{22}^2 c_\beta^2 - \text{Re}[M_{12}^2 e^{i \eta}] s_{2\beta} \,, \\
    m_{12}^2 e^{i\eta} &= \frac{1}{2} (M_{22}^2 - M_{11}^2 ) s_{2\beta} + \text{Re}[M_{12}^2 e^{i \eta}] c_{2\beta} + i \,\text{Im}[M_{12}^2 e^{i \eta}] \,,
\end{align}
\end{subequations}
where $\tan\beta = v_2/v_1$ with range $0 \leq \beta \leq \frac{\pi}{2}$, and $\eta$ is the phase accompanying $v_2$ in the general basis parameterization of the doublets in Eq. (\ref{generalvevs}). The relations between the quartic couplings are:
\begin{subequations}\label{quarticconversions}
\begin{align}
    \lambda_1 &= Z_1 c_\beta^4 + Z_2 s_\beta^4 + \frac{1}{2} Z_{345} s_{2\beta}^2 - 2 s_{2\beta} \left( \text{Re}[Z_6 e^{i\eta}] c_\beta^2 + \text{Re}[Z_7 e^{i\eta}] s_\beta^2 \right) \,, \\
    \lambda_2 &= Z_1 s_\beta^4 + Z_2 c_\beta^4 + \frac{1}{2} Z_{345} s_{2\beta}^2 + 2 s_{2\beta} \left( \text{Re}[Z_6 e^{i\eta}] s_\beta^2 + \text{Re}[Z_7 e^{i\eta}] c_\beta^2 \right) \,, \\
    \lambda_3 &= \frac{1}{4} \left( Z_1 + Z_2 - 2 Z_{345} \right) s_{2\beta}^2 + Z_3 + \text{Re}[(Z_6 - Z_7)e^{i\eta}] s_{2\beta} c_{2 \beta} \,, \\
    \lambda_4 &= \frac{1}{4} \left( Z_1 + Z_2 - 2 Z_{345} \right) s_{2\beta}^2 + Z_4 + \text{Re}[(Z_6 - Z_7)e^{i\eta}] s_{2\beta} c_{2 \beta} \,, \\
    \lambda_5 e^{2i \eta} &= \frac{1}{4} ( Z_1 + Z_2 - 2 Z_{345} ) s_{2\beta}^2 + \text{Re}[Z_5 e^{2i \eta}] + i\, \text{Im}[Z_5 e^{2i \eta}] c_{2\beta} \\ & \hspace{1cm} + \text{Re}[(Z_6 - Z_7)e^{i\eta}] s_{2\beta} c_{2 \beta} + i\, \text{Im}[(Z_6 - Z_7)e^{i\eta}] s_{2\beta} \,, \nonumber \\
    \lambda_6 e^{i \eta} &= \frac{1}{2} (Z_1 c_\beta^2 - Z_2 s_\beta^2 - Z_{345} c_{2\beta} - i\, \text{Im}[Z_5 e^{2i\eta}]) s_{2\beta} \\ & \hspace{1cm} + \text{Re}[Z_6 e^{i \eta}] c_\beta c_{3\beta} + i\,\text{Im}[Z_6 e^{i \eta}] c_\beta^2 + \text{Re}[Z_7 e^{i \eta}] s_{\beta} s_{3 \beta} + i\, \text{Im}[Z_7 e^{i \eta}] s_\beta^2 \,, \nonumber \\
    \lambda_7 e^{i \eta} & = \frac{1}{2} (Z_1 s_\beta^2 - Z_2 c_\beta^2 + Z_{345} c_{2\beta} + i\, \text{Im}[Z_5 e^{2i\eta}]) s_{2\beta} \\ & \hspace{1cm} + \text{Re}[Z_6 e^{i \eta}] s_\beta s_{3\beta} + i\,\text{Im}[Z_6 e^{i \eta}] s_\beta^2 + \text{Re}[Z_7 e^{i \eta}] c_{\beta} c_{3 \beta} + i\, \text{Im}[Z_7 e^{i \eta}] c_\beta^2 \,, \nonumber
\end{align}
\end{subequations}
where we have defined $Z_{345} \equiv (Z_3 + Z_4 + \text{Re}[Z_5 e^{2i \eta}])$.  For the reverse conversion from the Higgs basis to the general basis, one can perform the same series of identifications, but substituting $\lambda_i \leftrightarrow Z_i$ and $\beta \leftrightarrow - \beta$.

\section{Connection between \texorpdfstring{$\mathbf{\zeta}$}{zeta} and \texorpdfstring{$M_{H^\pm}^2$}{mHp}}\label{appA}

In \ccite{Ivanov:2006yq,Maniatis:2006fs}, it was mathematically demonstrated that the charged Higgs mass squared $m_{H^\pm}^2$ is related to the Lagrange multiplier $\zeta$ as:
\begin{equation}\label{claim}
	\zeta = \frac{M_{H^\pm}^2}{v^2} \,.
\end{equation}
We explain here why this is the case in the general 2HDM. Let the constrained function we want to extremize be:
$V = - M_\mu r^{\mu} + \frac{1}{2} \Lambda_{\mu \nu} r^\mu r^\nu - \frac{\zeta}{2} (r^\mu r_\mu- 2 c)$, with $M^\mu$, $r^\mu$, 
and $\Lambda_{\mu \nu}$ defined in Eqs.~(\ref{eq:Mmu}), (\ref{eq:rmu}) and (\ref{eq:Lambdamunu}), respectively. 
In the case presented in the main text, we have $c=0$, corresponding to the fact that the constraint is $r^\mu r_\mu = 0$ (i.e. no charge breaking minima are allowed). However, we will for the moment retain non-zero $c$ to see what happens when we allow the constraint to relax slightly. 

The claim is that $\zeta$ should be interpreted as the rate of change of the extremal value\footnote{In this appendix stars are used to denote quantities evaluated at an extremum. For convenience they are dropped in the body of the text.} $\overline{V}_*$ with respect to the constraint $c = \frac{1}{2} r^\mu r_\mu$, $\zeta = \frac{d \overline{V}_*}{dc}$; that is, it is the price one would pay to relax the constraint $r^{\mu}r_{\mu} = 0$. To see this, let $(r_*^\mu, \zeta_*)$ be the field configuration and Lagrange multiplier value which extremizes $\overline{V}$:
\begin{equation}
	\nabla \overline{V} \big|_{r_\star^\mu, \zeta_\star} = 0 \,,
\end{equation}
where $\nabla = \left( \frac{\partial}{\partial r^\mu}, \frac{\partial}{\partial \zeta} \right)$. Now if we elevate $c$ to a variable, then the extremal values $(r_*^\mu(c), \zeta_*(c))$ and minimum $\overline{V}_*(c)$ will vary with it. The total derivative of $V_*(r_*^\mu, \zeta_*, c)$ with respect to $c$ is:
\begin{equation}
	\frac{d \overline{V}_*}{dc} = \frac{\partial \overline{V}_*}{\partial r_*^\mu} \frac{d r_*^\mu}{dc} + \frac{\partial \overline{V}_*}{\partial \zeta_*} \frac{d \zeta_*}{dc} + \frac{\partial \overline{V}_*}{\partial c} = \frac{\partial \bar{V}_*}{\partial c} = \zeta_* \,,
\end{equation}
where we have used the fact that at the extremal point $\frac{\partial \overline{V}_*}{\partial r_*^\mu} = 0$ and $\frac{\partial \overline{V}_*}{\partial \zeta_*} = 0$. Thus it is indeed the case that the Lagrange multiplier tells us how the extremum (here the minimum value of the potential) changes when we relax the constraint:
\begin{equation}
	\zeta_* = \frac{d \overline{V}_*}{dc} \,.
\end{equation}
 
\noindent In our case when we allow for $r^\mu r_\mu >0$, the vacuum location shifts and becomes charge breaking, and the potential depth changes as well. How does the charged Higgs mass evaluated in the neutral minimum compare with the cost\footnote{We know this is a cost rather than a gain because it was shown in~\ccite{Ferreira:2004yd} that the potential value in a charge breaking minimum is always higher than that in the corresponding neutral minimum, $V_{CB} - V_N > 0$.} of increasing $r^{\mu}r_{\mu}$? To see this, we will work out the expression for $c$ in the case of a charge breaking minimum. \\
 
 We parameterize\footnote{The minimization conditions forbid both $w$ and the complex phase $\eta$ from being non-zero (i.e. we cannot have a vacuum which breaks both charge and CP), so we set $\eta=0$.} the expectation values of the doublets as:
\begin{equation}
	\expval{\Phi_1} = \frac{1}{\sqrt{2}} \begin{pmatrix} 0 \\ v_1 \end{pmatrix} \,, \,\,\, \expval{\Phi_2} = \frac{1}{\sqrt{2}} \begin{pmatrix} w \\ v_2 \end{pmatrix} \,.
\end{equation}
If we now expand $\frac{1}{2} r_*^\mu r_{*\, \mu}$ about this minimum and examine the fluctuations in the charged Higgs fields $\phi_1^\pm$ and $\phi_2^\pm$, after a bit of algebra one can show the quadratic part is:
\begin{equation}
\begin{split}
	c \equiv \frac{1}{2} r^\mu r_\mu & = 2 |\Phi_1^\dagger \Phi_1| |\Phi_2^\dagger \Phi_2| - 2 |\Phi_1^\dagger \Phi_2|^2 \\
	& = (w^2 + v_2)^2 \phi_1^+ \phi_1^- +  v_1^2 \phi_2^+ \phi_2^- -  w^2 \phi_1^+ \Phi_1^- -  v_1 v_2 (\phi_1^+ \phi_2^- + \phi_2^+ \phi_2^-)\\
	& = \begin{pmatrix} \phi_1^-, & \phi_2^- \end{pmatrix} \begin{pmatrix} v_2^2 & - v_1 v_2 \\ - v_1 v_2 & v_1^2 \end{pmatrix} \begin{pmatrix} \phi_1^+ \\ \phi_2^+ \end{pmatrix}  \,.
\end{split}
\end{equation}
Similarly, in expanding about the minimum\footnote{Note that $\overline{V}_* = V_*$, since by design the constraint vanishes on the solution $ c - \frac{1}{2} r_*^\mu r_{* \mu} = 0$.} of the potential $V_*$, the quadratic part for the charged Higgs fields is:
\begin{equation}
	V_* \supset \frac{m_{H^\pm}^2}{v^2} \begin{pmatrix} \phi_1^-, & \phi_2^- \end{pmatrix} \begin{pmatrix} v_2^2 & - v_1 v_2 \\ - v_1 v_2 & v_1^2 \end{pmatrix} \begin{pmatrix} \phi_1^+ \\ \phi_2^+ \end{pmatrix} \,,
\end{equation}
with $H^\pm$ related to $\phi^\pm$ as:
\begin{equation}
	H^\pm = \frac{v_2}{v} \phi_1^\pm - \frac{v_1}{v} \phi_2^\pm \,.
\end{equation}
Thus we immediately see that 
\begin{equation}
	\frac{d V_*}{dc} = \frac{M_{H^\pm}^2}{v^2} \,.
\end{equation}
Since $\zeta_* = \frac{d V_*}{dc}$, we verify the claim\footnote{Technically the $\zeta$ of Eq. (\ref{claim}) is $\zeta_*$, since it is evaluated in a vacuum field configuration (recall that the star notation denoted a field configuration extremizing the potential). In the rest of the paper we drop the star notation.} of Eq.~(\ref{claim}). Thus, it will generically be the case that the Lagrange multiplier accompanying the constraint $r^\mu r_\mu = 0$ is proportional to the charged Higgs mass squared in the neutral vacuum.

%% file: bibliography.bib
@article{Pilaftsis:1999qt,
    author = "Pilaftsis, Apostolos and Wagner, Carlos E. M.",
    title = "{Higgs bosons in the minimal supersymmetric standard model with explicit CP violation}",
    eprint = "hep-ph/9902371",
    archivePrefix = "arXiv",
    reportNumber = "CERN-TH-99-34",
    doi = "10.1016/S0550-3213(99)00261-8",
    journal = "Nucl. Phys. B",
    volume = "553",
    pages = "3--42",
    year = "1999"
}

@article{Carena:1995bx,
    author = "Carena, Marcela and Espinosa, J. R. and Quiros, M. and Wagner, C. E. M.",
    title = "{Analytical expressions for radiatively corrected Higgs masses and couplings in the MSSM}",
    eprint = "hep-ph/9504316",
    archivePrefix = "arXiv",
    reportNumber = "CERN-TH-95-45, CERN-TH-95-045, DESY-95-038, IEM-FT-103-95",
    doi = "10.1016/0370-2693(95)00694-G",
    journal = "Phys. Lett. B",
    volume = "355",
    pages = "209--221",
    year = "1995"
}

@article{Englert:1964et,
    author = "Englert, F. and Brout, R.",
    editor = "Taylor, J. C.",
    title = "{Broken Symmetry and the Mass of Gauge Vector Mesons}",
    doi = "10.1103/PhysRevLett.13.321",
    journal = "Phys. Rev. Lett.",
    volume = "13",
    pages = "321--323",
    year = "1964"
}

@article{Higgs:1964pj,
    author = "Higgs, Peter W.",
    editor = "Taylor, J. C.",
    title = "{Broken Symmetries and the Masses of Gauge Bosons}",
    doi = "10.1103/PhysRevLett.13.508",
    journal = "Phys. Rev. Lett.",
    volume = "13",
    pages = "508--509",
    year = "1964"
}

@article{Guralnik:1964eu,
    author = "Guralnik, G. S. and Hagen, C. R. and Kibble, T. W. B.",
    editor = "Taylor, J. C.",
    title = "{Global Conservation Laws and Massless Particles}",
    doi = "10.1103/PhysRevLett.13.585",
    journal = "Phys. Rev. Lett.",
    volume = "13",
    pages = "585--587",
    year = "1964"
}

@article{CMS:2018uag,
    author = "Sirunyan, Albert M and others",
    collaboration = "CMS",
    title = "{Combined measurements of Higgs boson couplings in proton\textendash{}proton collisions at $\sqrt{s}=13\,\text {Te}\text {V} $}",
    eprint = "1809.10733",
    archivePrefix = "arXiv",
    primaryClass = "hep-ex",
    reportNumber = "CMS-HIG-17-031, CERN-EP-2018-263",
    doi = "10.1140/epjc/s10052-019-6909-y",
    journal = "Eur. Phys. J. C",
    volume = "79",
    number = "5",
    pages = "421",
    year = "2019"
}

@article{ATLAS:2019nkf,
    author = "Aad, Georges and others",
    collaboration = "ATLAS",
    title = "{Combined measurements of Higgs boson production and decay using up to $80$ fb$^{-1}$ of proton-proton collision data at $\sqrt{s}=$ 13 TeV collected with the ATLAS experiment}",
    eprint = "1909.02845",
    archivePrefix = "arXiv",
    primaryClass = "hep-ex",
    reportNumber = "CERN-EP-2019-097",
    doi = "10.1103/PhysRevD.101.012002",
    journal = "Phys. Rev. D",
    volume = "101",
    number = "1",
    pages = "012002",
    year = "2020"
}

@article{ACME:2018yjb,
    author = "Andreev, V. and others",
    collaboration = "ACME",
    title = "{Improved limit on the electric dipole moment of the electron}",
    doi = "10.1038/s41586-018-0599-8",
    journal = "Nature",
    volume = "562",
    number = "7727",
    pages = "355--360",
    year = "2018"
}

@article{Low:2002ws,
    author = "Low, Ian and Skiba, Witold and Tucker-Smith, David",
    title = "{Little Higgses from an antisymmetric condensate}",
    eprint = "hep-ph/0207243",
    archivePrefix = "arXiv",
    reportNumber = "HUTP-02-A033, MIT-CTP-3291",
    doi = "10.1103/PhysRevD.66.072001",
    journal = "Phys. Rev. D",
    volume = "66",
    pages = "072001",
    year = "2002"
}

@article{Gunion:2002zf,
    author = "Gunion, John F. and Haber, Howard E.",
    title = "{The CP conserving two Higgs doublet model: The Approach to the decoupling limit}",
    eprint = "hep-ph/0207010",
    archivePrefix = "arXiv",
    reportNumber = "SCIPP-02-10",
    doi = "10.1103/PhysRevD.67.075019",
    journal = "Phys. Rev. D",
    volume = "67",
    pages = "075019",
    year = "2003"
}

@article{Carena:2013ooa,
    author = "Carena, Marcela and Low, Ian and Shah, Nausheen R. and Wagner, Carlos E. M.",
    title = "{Impersonating the Standard Model Higgs Boson: Alignment without Decoupling}",
    eprint = "1310.2248",
    archivePrefix = "arXiv",
    primaryClass = "hep-ph",
    reportNumber = "ANL-HEP-PR-13-50, EFI-13-27, FERMILAB-PUB-13-455-T, MCTP-13-31",
    doi = "10.1007/JHEP04(2014)015",
    journal = "JHEP",
    volume = "04",
    pages = "015",
    year = "2014"
}

@article{BhupalDev:2014bir,
    author = "Bhupal Dev, P. S. and Pilaftsis, Apostolos",
    title = "{Maximally Symmetric Two Higgs Doublet Model with Natural Standard Model Alignment}",
    eprint = "1408.3405",
    archivePrefix = "arXiv",
    primaryClass = "hep-ph",
    reportNumber = "MAN-HEP-2014-10, CERN-PH-TH-2014-150",
    doi = "10.1007/JHEP12(2014)024",
    journal = "JHEP",
    volume = "12",
    pages = "024",
    year = "2014",
    note = "[Erratum: JHEP 11, 147 (2015)]"
}

@article{Carena:2014nza,
    author = "Carena, Marcela and Haber, Howard E. and Low, Ian and Shah, Nausheen R. and Wagner, Carlos E. M.",
    title = "{Complementarity between Nonstandard Higgs Boson Searches and Precision Higgs Boson Measurements in the MSSM}",
    eprint = "1410.4969",
    archivePrefix = "arXiv",
    primaryClass = "hep-ph",
    reportNumber = "EFI-14-36, FERMILAB-PUB-14-392-T, MCTP-14-37, SCIPP-14-16",
    doi = "10.1103/PhysRevD.91.035003",
    journal = "Phys. Rev. D",
    volume = "91",
    number = "3",
    pages = "035003",
    year = "2015"
}

@article{Carena:2015moc,
    author = "Carena, Marcela and Haber, Howard E. and Low, Ian and Shah, Nausheen R. and Wagner, Carlos E. M.",
    title = "{Alignment limit of the NMSSM Higgs sector}",
    eprint = "1510.09137",
    archivePrefix = "arXiv",
    primaryClass = "hep-ph",
    reportNumber = "FERMILAB-PUB-15-407-T, EFI-15-32, MCTP-15-15, SCIPP-15-12, WSU-HEP-1505",
    doi = "10.1103/PhysRevD.93.035013",
    journal = "Phys. Rev. D",
    volume = "93",
    number = "3",
    pages = "035013",
    year = "2016"
}

@article{Bernon:2015qea,
    author = "Bernon, J\'er\'emy and Gunion, John F. and Haber, Howard E. and Jiang, Yun and Kraml, Sabine",
    title = "{Scrutinizing the alignment limit in two-Higgs-doublet models: m$_h$=125  GeV}",
    eprint = "1507.00933",
    archivePrefix = "arXiv",
    primaryClass = "hep-ph",
    doi = "10.1103/PhysRevD.92.075004",
    journal = "Phys. Rev. D",
    volume = "92",
    number = "7",
    pages = "075004",
    year = "2015"
}

@article{ParticleDataGroup:2018ovx,
    author = "Tanabashi, M. and others",
    collaboration = "Particle Data Group",
    title = "{Review of Particle Physics}",
    doi = "10.1103/PhysRevD.98.030001",
    journal = "Phys. Rev. D",
    volume = "98",
    number = "3",
    pages = "030001",
    year = "2018"
}

@article{Coyle:2018ydo,
    author = "Coyle, Nina M. and Li, Bing and Wagner, Carlos E. M.",
    title = "{Wrong sign bottom Yukawa coupling in low energy supersymmetry}",
    eprint = "1802.09122",
    archivePrefix = "arXiv",
    primaryClass = "hep-ph",
    doi = "10.1103/PhysRevD.97.115028",
    journal = "Phys. Rev. D",
    volume = "97",
    number = "11",
    pages = "115028",
    year = "2018"
}

@article{Bento:2022vsb,
    author = "Bento, Miguel P. and Rom\~ao, Jorge C. and Silva, Jo\~ao P.",
    title = "{Unitarity bounds for all symmetry-constrained 3HDMs}",
    eprint = "2204.13130",
    archivePrefix = "arXiv",
    primaryClass = "hep-ph",
    month = "4",
    year = "2022"
}

@article{Ivanov:2015nea,
    author = "Ivanov, I. P. and Silva, Joao P.",
    title = "{Tree-level metastability bounds for the most general two Higgs doublet model}",
    eprint = "1507.05100",
    archivePrefix = "arXiv",
    primaryClass = "hep-ph",
    reportNumber = "CFTP-15-007",
    doi = "10.1103/PhysRevD.92.055017",
    journal = "Phys. Rev. D",
    volume = "92",
    number = "5",
    pages = "055017",
    year = "2015"
}

@article{doi:10.1137/0915035,
	author = {Ulrich, Gary and Watson, Layne T.},
	doi = {10.1137/0915035},
	eprint = {https://doi.org/10.1137/0915035},
	journal = {SIAM Journal on Scientific Computing},
	number = {3},
	pages = {528-544},
	title = {Positivity Conditions for Quartic Polynomials},
	url = {https://doi.org/10.1137/0915035},
	volume = {15},
	year = {1994}
}

@article{Kanemura:2015ska,
    author = "Kanemura, Shinya and Yagyu, Kei",
    title = "{Unitarity bound in the most general two Higgs doublet model}",
    eprint = "1509.06060",
    archivePrefix = "arXiv",
    primaryClass = "hep-ph",
    reportNumber = "UT-HET-107",
    doi = "10.1016/j.physletb.2015.10.047",
    journal = "Phys. Lett. B",
    volume = "751",
    pages = "289--296",
    year = "2015"
}

@article{Barroso:2013awa,
    author = "Barroso, A. and Ferreira, P. M. and Ivanov, I. P. and Santos, Rui",
    title = "{Metastability bounds on the two Higgs doublet model}",
    eprint = "1303.5098",
    archivePrefix = "arXiv",
    primaryClass = "hep-ph",
    doi = "10.1007/JHEP06(2013)045",
    journal = "JHEP",
    volume = "06",
    pages = "045",
    year = "2013"
}

@article{Gershgorin:1931,
    author = "Gershgorin, S.",
    title = "{\"{U}ber die Abgrenzung der Eigenwerte einer Matrix}",
    journal = "Bulletin de l'Academie des Sciences de l'URSS",
    volume = "6",
    number = "3",
    pages = "749-754",
    year = "1931"
}

@article{Branco:2011iw,
    author = "Branco, G. C. and Ferreira, P. M. and Lavoura, L. and Rebelo, M. N. and Sher, Marc and Silva, Joao P.",
    title = "{Theory and phenomenology of two-Higgs-doublet models}",
    eprint = "1106.0034",
    archivePrefix = "arXiv",
    primaryClass = "hep-ph",
    doi = "10.1016/j.physrep.2012.02.002",
    journal = "Phys. Rept.",
    volume = "516",
    pages = "1--102",
    year = "2012"
}

@article{Lee:2015uza,
    author = "Lee, Gabriel and Wagner, Carlos E. M.",
    title = "{Higgs bosons in heavy supersymmetry with an intermediate m$_A$}",
    eprint = "1508.00576",
    archivePrefix = "arXiv",
    primaryClass = "hep-ph",
    reportNumber = "EFI-PREPRINT-15-24",
    doi = "10.1103/PhysRevD.92.075032",
    journal = "Phys. Rev. D",
    volume = "92",
    number = "7",
    pages = "075032",
    year = "2015"
}

@article{Bahl:2018jom,
    author = "Bahl, Henning and Hollik, Wolfgang",
    title = "{Precise prediction of the MSSM Higgs boson masses for low M$_{A}$}",
    eprint = "1805.00867",
    archivePrefix = "arXiv",
    primaryClass = "hep-ph",
    doi = "10.1007/JHEP07(2018)182",
    journal = "JHEP",
    volume = "07",
    pages = "182",
    year = "2018"
}

@article{Bahl:2020mjy,
    author = "Bahl, Henning and Murphy, Nick and Rzehak, Heidi",
    title = "{Hybrid calculation of the MSSM Higgs boson masses using the complex THDM as EFT}",
    eprint = "2010.04711",
    archivePrefix = "arXiv",
    primaryClass = "hep-ph",
    doi = "10.1140/epjc/s10052-021-08939-7",
    journal = "Eur. Phys. J. C",
    volume = "81",
    number = "2",
    pages = "128",
    year = "2021"
}

@article{Bahl:2020jaq,
    author = "Bahl, Henning and Sobolev, Ivan",
    title = "{Two-loop matching of renormalizable operators: general considerations and applications}",
    eprint = "2010.01989",
    archivePrefix = "arXiv",
    primaryClass = "hep-ph",
    reportNumber = "DESY-20-153, DESY 20-153",
    doi = "10.1007/JHEP03(2021)286",
    journal = "JHEP",
    volume = "03",
    pages = "286",
    year = "2021"
}

@article{Carena:2015uoe,
    author = "Carena, M. and Ellis, J. and Lee, J. S. and Pilaftsis, A. and Wagner, C. E. M.",
    title = "{CP Violation in Heavy MSSM Higgs Scenarios}",
    eprint = "1512.00437",
    archivePrefix = "arXiv",
    primaryClass = "hep-ph",
    reportNumber = "CNU-HEP-15-07, FERMILAB-PUB-15-508-T, EFI-15-36, MAN-HEP-2015-19, LCTS-2015-37, CERN-PH-TH-2015-252, KCL-PH-TH-2015-49",
    doi = "10.1007/JHEP02(2016)123",
    journal = "JHEP",
    volume = "02",
    pages = "123",
    year = "2016"
}

@article{Haber:1993an,
    author = "Haber, Howard E. and Hempfling, Ralf",
    title = "{The Renormalization group improved Higgs sector of the minimal supersymmetric model}",
    eprint = "hep-ph/9307201",
    archivePrefix = "arXiv",
    reportNumber = "SCIPP-91-33-REV, SCIPP-91-33",
    doi = "10.1103/PhysRevD.48.4280",
    journal = "Phys. Rev. D",
    volume = "48",
    pages = "4280--4309",
    year = "1993"
}

@article{Murphy:2019qpm,
    author = "Murphy, Nick and Rzehak, Heidi",
    title = "{Higgs-Boson Masses and Mixings in the MSSM with CP Violation and Heavy SUSY Particles}",
    eprint = "1909.00726",
    archivePrefix = "arXiv",
    primaryClass = "hep-ph",
    month = "9",
    year = "2019"
}

@article{Kanemura:1993hm,
    author = "Kanemura, Shinya and Kubota, Takahiro and Takasugi, Eiichi",
    title = "{Lee-Quigg-Thacker bounds for Higgs boson masses in a two doublet model}",
    eprint = "hep-ph/9303263",
    archivePrefix = "arXiv",
    reportNumber = "OS-GE-32-93",
    doi = "10.1016/0370-2693(93)91205-2",
    journal = "Phys. Lett. B",
    volume = "313",
    pages = "155--160",
    year = "1993"
}

@article{Horejsi:2005da,
    author = "Horejsi, J. and Kladiva, M.",
    title = "{Tree-unitarity bounds for THDM Higgs masses revisited}",
    eprint = "hep-ph/0510154",
    archivePrefix = "arXiv",
    doi = "10.1140/epjc/s2006-02472-3",
    journal = "Eur. Phys. J. C",
    volume = "46",
    pages = "81--91",
    year = "2006"
}

@article{Ginzburg:2005dt,
    author = "Ginzburg, I. F. and Ivanov, I. P.",
    title = "{Tree-level unitarity constraints in the most general 2HDM}",
    eprint = "hep-ph/0508020",
    archivePrefix = "arXiv",
    doi = "10.1103/PhysRevD.72.115010",
    journal = "Phys. Rev. D",
    volume = "72",
    pages = "115010",
    year = "2005"
}

@article{Akeroyd:2000wc,
    author = "Akeroyd, Andrew G. and Arhrib, Abdesslam and Naimi, El-Mokhtar",
    title = "{Note on tree level unitarity in the general two Higgs doublet model}",
    eprint = "hep-ph/0006035",
    archivePrefix = "arXiv",
    reportNumber = "UFR-HEP-00-06, KEK-TH-00-699, KEK-TH-699",
    doi = "10.1016/S0370-2693(00)00962-X",
    journal = "Phys. Lett. B",
    volume = "490",
    pages = "119--124",
    year = "2000"
}

@article{Barroso:2007rr,
    author = "Barroso, A. and Ferreira, P. M. and Santos, R.",
    title = "{Neutral minima in two-Higgs doublet models}",
    eprint = "hep-ph/0702098",
    archivePrefix = "arXiv",
    doi = "10.1016/j.physletb.2007.07.010",
    journal = "Phys. Lett. B",
    volume = "652",
    pages = "181--193",
    year = "2007"
}

@article{Ivanov:2007de,
    author = "Ivanov, Igor P.",
    title = "{Minkowski space structure of the Higgs potential in 2HDM. II. Minima, symmetries, and topology}",
    eprint = "0710.3490",
    archivePrefix = "arXiv",
    primaryClass = "hep-ph",
    doi = "10.1103/PhysRevD.77.015017",
    journal = "Phys. Rev. D",
    volume = "77",
    pages = "015017",
    year = "2008"
}

@article{Barroso:2005sm,
    author = "Barroso, A. and Ferreira, P. M. and Santos, R.",
    title = "{Charge and CP symmetry breaking in two Higgs doublet models}",
    eprint = "hep-ph/0507224",
    archivePrefix = "arXiv",
    doi = "10.1016/j.physletb.2005.11.031",
    journal = "Phys. Lett. B",
    volume = "632",
    pages = "684--687",
    year = "2006"
}

@article{Maniatis:2006fs,
    author = "Maniatis, M. and von Manteuffel, A. and Nachtmann, O. and Nagel, F.",
    title = "{Stability and symmetry breaking in the general two-Higgs-doublet model}",
    eprint = "hep-ph/0605184",
    archivePrefix = "arXiv",
    reportNumber = "HD-THEP-06-07",
    doi = "10.1140/epjc/s10052-006-0016-6",
    journal = "Eur. Phys. J. C",
    volume = "48",
    pages = "805--823",
    year = "2006"
}

@article{Ivanov:2006yq,
    author = "Ivanov, I. P.",
    title = "{Minkowski space structure of the Higgs potential in 2HDM}",
    eprint = "hep-ph/0609018",
    archivePrefix = "arXiv",
    doi = "10.1103/PhysRevD.75.035001",
    journal = "Phys. Rev. D",
    volume = "75",
    pages = "035001",
    year = "2007",
    note = "[Erratum: Phys.Rev.D 76, 039902 (2007)]"
}

@article{Ferreira:2004yd,
    author = "Ferreira, P. M. and Santos, R. and Barroso, A.",
    title = "{Stability of the tree-level vacuum in two Higgs doublet models against charge or CP spontaneous violation}",
    eprint = "hep-ph/0406231",
    archivePrefix = "arXiv",
    doi = "10.1016/j.physletb.2004.10.022",
    journal = "Phys. Lett. B",
    volume = "603",
    pages = "219--229",
    year = "2004",
    note = "[Erratum: Phys.Lett.B 629, 114--114 (2005)]"
}

@article{Ferreira:2009jb,
    author = "Ferreira, P. M. and Jones, D. R. T.",
    title = "{Bounds on scalar masses in two Higgs doublet models}",
    eprint = "0903.2856",
    archivePrefix = "arXiv",
    primaryClass = "hep-ph",
    doi = "10.1088/1126-6708/2009/08/069",
    journal = "JHEP",
    volume = "08",
    pages = "069",
    year = "2009"
}

@article{Klimenko:1984qx,
    author = "Klimenko, K. G.",
    title = "{On Necessary and Sufficient Conditions for Some Higgs Potentials to Be Bounded From Below}",
    reportNumber = "IFVE-84-43",
    doi = "10.1007/BF01034825",
    journal = "Theor. Math. Phys.",
    volume = "62",
    pages = "58--65",
    year = "1985"
}

@article{Bahl:2019ago,
    author = "Bahl, Henning and Liebler, Stefan and Stefaniak, Tim",
    title = "{MSSM Higgs benchmark scenarios for Run 2 and beyond: the low $\tan \beta $ region}",
    eprint = "1901.05933",
    archivePrefix = "arXiv",
    primaryClass = "hep-ph",
    reportNumber = "DESY 18-219, DESY-18-219, KA-TP-01-2019",
    doi = "10.1140/epjc/s10052-019-6770-z",
    journal = "Eur. Phys. J. C",
    volume = "79",
    number = "3",
    pages = "279",
    year = "2019"
}

@article{Low:2020iua,
    author = "Low, Ian and Shah, Nausheen R. and Wang, Xiao-Ping",
    title = "{Higgs alignment and novel CP-violating observables in two-Higgs-doublet models}",
    eprint = "2012.00773",
    archivePrefix = "arXiv",
    primaryClass = "hep-ph",
    reportNumber = "WSU-HEP-2005",
    doi = "10.1103/PhysRevD.105.035009",
    journal = "Phys. Rev. D",
    volume = "105",
    number = "3",
    pages = "035009",
    year = "2022"
}

@article{Jurciukonis:2018skr,
    author = "Jur\v{c}iukonis, D. and Lavoura, L.",
    title = "{The three- and four-Higgs couplings in the general two-Higgs-doublet model}",
    eprint = "1807.04244",
    archivePrefix = "arXiv",
    primaryClass = "hep-ph",
    doi = "10.1007/JHEP12(2018)004",
    journal = "JHEP",
    volume = "12",
    pages = "004",
    year = "2018"
}

@article{Lavoura:1994fv,
    author = "Lavoura, L. and Silva, Joao P.",
    title = "{Fundamental CP violating quantities in a SU(2) x U(1) model with many Higgs doublets}",
    eprint = "hep-ph/9404276",
    archivePrefix = "arXiv",
    doi = "10.1103/PhysRevD.50.4619",
    journal = "Phys. Rev. D",
    volume = "50",
    pages = "4619--4624",
    year = "1994"
}

@article{Haber:2015pua,
    author = "Haber, Howard E. and St\r{a}l, Oscar",
    title = "{New LHC benchmarks for the $\mathcal{CP}$ -conserving two-Higgs-doublet model}",
    eprint = "1507.04281",
    archivePrefix = "arXiv",
    primaryClass = "hep-ph",
    reportNumber = "SCIPP-15-10",
    doi = "10.1140/epjc/s10052-015-3697-x",
    journal = "Eur. Phys. J. C",
    volume = "75",
    number = "10",
    pages = "491",
    year = "2015",
    note = "[Erratum: Eur.Phys.J.C 76, 312 (2016)]"
}

@phdthesis{Garren,
    author = "Garren, Kenneth Ross",
    title = "{Bounds for the Eigenvalues of a Matrix}",
    journal = "Dissertations, Theses, and Masters Projects",
    school = "William and Mary",
    year = "1965",
    doi = "10.21220/s2-p2tn-7046"
}

@article{Haber:2022gsn,
    author = "Haber, Howard E. and Keus, Venus and Santos, Rui",
    title = "{P-even, CP-violating Signals in Scalar-Mediated Processes}",
    eprint = "2206.09643",
    archivePrefix = "arXiv",
    primaryClass = "hep-ph",
    reportNumber = "DIAS-STP-22-02, SCIPP-22/01",
    month = "6",
    year = "2022"
}

@article{Boto:2020wyf,
    author = "Boto, Rafael and Fernandes, Tiago V. and Haber, Howard E. and Rom\~ao, Jorge C. and Silva, Jo\~ao P.",
    title = "{Basis-independent treatment of the complex 2HDM}",
    eprint = "2001.01430",
    archivePrefix = "arXiv",
    primaryClass = "hep-ph",
    reportNumber = "CFTP/19-032 and SCIPP-19/01, SCIPP-19/01",
    doi = "10.1103/PhysRevD.101.055023",
    journal = "Phys. Rev. D",
    volume = "101",
    number = "5",
    pages = "055023",
    year = "2020"
}

@article{Davidson:2005cw,
    author = "Davidson, Sacha and Haber, Howard E.",
    title = "{Basis-independent methods for the two-Higgs-doublet model}",
    eprint = "hep-ph/0504050",
    archivePrefix = "arXiv",
    reportNumber = "IPPP-03-23, DCPT-03-46, SCIPP-04-15",
    doi = "10.1103/PhysRevD.72.099902",
    journal = "Phys. Rev. D",
    volume = "72",
    pages = "035004",
    year = "2005",
    note = "[Erratum: Phys.Rev.D 72, 099902 (2005)]"
}

@article{Song:2022gsz,
    author = "Song, Yisheng",
    title = "{Co-positivity of tensors and Stability conditions of two Higgs potentials}",
    eprint = "2203.11462",
    archivePrefix = "arXiv",
    primaryClass = "hep-ph",
    month = "3",
    year = "2022"
}

@article{Pich:2009sp,
    author = "Pich, Antonio and Tuzon, Paula",
    title = "{Yukawa Alignment in the Two-Higgs-Doublet Model}",
    eprint = "0908.1554",
    archivePrefix = "arXiv",
    primaryClass = "hep-ph",
    reportNumber = "IFIC-09-36, FTUV-09-0810",
    doi = "10.1103/PhysRevD.80.091702",
    journal = "Phys. Rev. D",
    volume = "80",
    pages = "091702",
    year = "2009"
}

@article{Altmannshofer:2020shb,
    author = "Altmannshofer, Wolfgang and Gori, Stefania and Hamer, Nick and Patel, Hiren H.",
    title = "{Electron EDM in the complex two-Higgs doublet model}",
    eprint = "2009.01258",
    archivePrefix = "arXiv",
    primaryClass = "hep-ph",
    doi = "10.1103/PhysRevD.102.115042",
    journal = "Phys. Rev. D",
    volume = "102",
    number = "11",
    pages = "115042",
    year = "2020"
}
